\documentclass[modern]{aastex7}
\usepackage{natbib}      
\usepackage{graphicx}    
\usepackage{amsmath}
\usepackage{amssymb}
\usepackage{accents}
\usepackage{bm} 
\usepackage{sidecap}
\sidecaptionvpos{figure}{t}

\usepackage{comment}

\usepackage{fancyhdr}
\usepackage{epsfig}
\usepackage{float}
\usepackage{sidecap}\sidecaptionvpos{figure}{t}

\usepackage{url}

\usepackage{verbatim}     

\usepackage{graphicx}    
\usepackage{stmaryrd} 
\usepackage{calc, layouts}

\hypersetup{linkcolor=red,citecolor=cyan,filecolor=cyan,urlcolor=blue}

\received{2024 October 2}
\revised{2025 May 8}
\accepted{2025 May 12}

\submitjournal{ApJ}

\shorttitle{Magnetogram-matching Biot\textendash{}Savart Law and Decomposition}
\shortauthors{Titov et al.}

\usepackage{color}
\definecolor{Light}{gray}{.50}
\definecolor{Dark}{gray}{.20}

\definecolor{dark-red}{rgb}{0.8,0,0}
\definecolor{dark-green}{rgb}{0,0.6,0}
\definecolor{dark-blue}{rgb}{0,0,0.8}
\definecolor{dark-magenta}{rgb}{0.8,0,0.8}
\definecolor{dark-purple}{rgb}{0.45,0.2,0.65}
\definecolor{orange}{rgb}{1.0,0.6,0}
\usepackage{soul}

\definecolor{Light}{gray}{.50}
\definecolor{Dark}{gray}{.20}
\definecolor{dark-red}{rgb}{0.8,0,0}
\definecolor{dark-green}{rgb}{0,0.6,0}
\definecolor{dark-blue}{rgb}{0,0,0.8}
\definecolor{dark-magenta}{rgb}{0.8,0,0.8}
\definecolor{orange}{rgb}{1.0,0.6,0}
\DeclareMathOperator{\slog}{slog}

\newcommand{\Cs}{{\mathcal C}^{*}}
\newcommand{\Css}{{\mathcal C}^{**}}
\newcommand{\Cc}{{\mathcal C}}
\newcommand{\Ss}{{\mathcal S}^{*}}
\newcommand{\Sss}{{\mathcal S}^{**}}

\newcommand{\Bv}{{\bm B}}
\newcommand{\BPlt}{ \Bv^{<}_{\mathrm{P}} }
\newcommand{\BPgt}{ \Bv^{>}_{\mathrm{P}} }
\newcommand{\Jv}{ \bm{J} }

\newcommand{\St}{{\skew4\tilde{S}}}
\newcommand{\Bpot}{\Bv_{\mathrm{pot}}}
\newcommand{\Btld}{{\skew4\tilde{\Bv}}}
\newcommand{\BvT}{\Bv_{T}}
\newcommand{\BvSt}{\Bv_{\St}}
\newcommand{\Rv}{ \bm{\mathcal{R}} }
\newcommand{\Rvs}{ \bm{\mathcal{R}}_{*} }
\newcommand{\Rvp}{ \bm{\mathcal{R}}^{\prime} }
\newcommand{\bcd}
               { \bm{\cdot} }
\newcommand{\bmt}
               { \bm{\times} }
\newcommand{\mv}{\bm{m}}
\newcommand{\mvh}{\skew3\hat{\mv}}

\newcommand{\ex}{\skew3\hat{\bm{x}} }

\newcommand{\ez}{\skew3\hat{\bm{z}} }
\newcommand{\dx}{\mathrm{d}x}

\newcommand{\dz}{\mathrm{d}z}

\newcommand{\et}{\skew3\hat{\bm{\theta}}}
\newcommand{\ep}{\skew4\hat{\bm{\phi}}}

\newcommand{\nt}{\bm{\nabla}_\mathrm{t}}
\newcommand{\Rvu}{\skew3\hat{\bm{R}}}

\newcommand{\xv}{ \bm{x} }
\newcommand{\rv}{ \bm{r} }

\newcommand{\rvs}{ \bm{r}_{*} }
\newcommand{\rs}{r_{*}}

\newcommand{\est}[1]
{
\raisebox{-3pt}{$\Bigr\vert$}_{\raisebox{4pt}{$_{#1}$}}
}
\newcommand{\RBSL}{{\footnotesize R}BS{\footnotesize L}}
\newcommand{\BSL}{BS{\footnotesize L}}
\newcommand{\MBSL}{{\footnotesize M}BS{\footnotesize L}}
\newcommand{\Bvdec}{
\Bv = \Bpot + \BvT + \BvSt}
\newcommand{\BvmBpot}{
\Bv - \Bpot}
\begin{document}


\title{Magnetogram-matching Biot\textendash{}Savart Law and Decomposition of Vector Magnetograms} 

\author[orcid=0000-0001-7053-4081,sname='Titov']{Viacheslav S. Titov}
\affiliation{Predictive Science Inc., 9990 Mesa Rim Road, Suite 170, San Diego, CA 92121, USA} 
\email[show]{titovv@predsci.com}

\author[orcid=0000-0003-1759-4354,sname='Downs']{Cooper Downs}
\affiliation{Predictive Science Inc., 9990 Mesa Rim Road, Suite 170, San Diego, CA 92121, USA} 
\email{cdowns@predsci.com}

\author[orcid=0000-0003-3843-3242,sname='T\"{o}r\"{o}k']{Tibor T\"{o}r\"{o}k}
\affiliation{Predictive Science Inc., 9990 Mesa Rim Road, Suite 170, San Diego, CA 92121, USA} 
\email{tibor@predsci.com}

\author[orcid=0000-0003-1662-3328,sname='Linker']{ Jon A. Linker}
\affiliation{Predictive Science Inc., 9990 Mesa Rim Road, Suite 170, San Diego, CA 92121, USA} 
\email{linkerj@predsci.com}

\author[orcid=0009-0006-4049-836X,sname='Prazak']{Michael Prazak}
\affiliation{Department of Physics, Montana State University, Bozeman, MT 59717, USA} 
\email{michaelprazak@montana.edu}

\author[0000-0002-2797-744X,sname='Qiu']{ Jiong A. Qiu}
\affiliation{Department of Physics, Montana State University, Bozeman, MT 59717, USA} 
\email{qiu@montana.edu}

\begin{abstract}

We generalize a magnetogram-matching Biot\textendash{}Savart law (\BSL) from planar to spherical geometry.
For a given coronal current density $\Jv$, this law determines the magnetic field $\Btld$ whose radial component vanishes at the surface.
The superposition of $\Btld$ with a potential field defined by a given surface radial field, $B_r$, provides the entire configuration where $B_r$ remains unchanged by the currents. 
Using this approach, we (1) upgrade our regularized \BSL{s} for constructing coronal magnetic flux ropes (MFRs) and (2) propose a new method for decomposing a measured photospheric magnetic field as $\Bvdec$, where the potential, $\Bpot$, toroidal, $\BvT$, and poloidal, $\BvSt$, fields are determined by $B_r$, $J_r$, and the surface divergence of $\BvmBpot$, respectively, all derived from magnetic data.
Our $\BvT$ is identical to the one in the alternative Gaussian decomposition by P. W. Schuck et al. (2022), while $\Bpot$ and $\BvSt$ are different from their poloidal fields $\BPlt$ and $\BPgt$, which are {\it potential} 
in the infinitesimal proximity to the upper and lower side of the surface, respectively.
In contrast, our $\BvSt$
has no  such constraints and, as $\Bpot$  and $\BvT$, refers to the {\it same} upper side of the surface.
In spite of these differences,  for a continuous  $\Jv$ distribution across the surface, $\Bpot$ and $\BvSt$ are linear combinations of $\BPlt$ and $\BPgt$.
We demonstrate that, similar to the Gaussian method, our decomposition
allows one to identify the footprints and projected surface-location
of MFRs in the solar corona, as well as the direction and connectivity of their currents.

\end{abstract}

\keywords{Sun: coronal mass ejections (CMEs)---Sun: flares---Sun: magnetic fields} 

\section{Introduction
	\label{s:intro}}

Magnetic fields and electric currents play a key role in many dynamic processes in the solar atmosphere \citep[e.g.,][]{priest14}.
Among them, solar eruptions are the most energetic and probably most spectacular phenomena, often producing coronal mass ejections \citep[CMEs; e.g.,][]{webb12} that propagate far beyond the corona.
These gigantic ejections of magnetized plasma can cause dangerous streams of accelerated particles penetrating interplanetary space \citep[e.g.,][]{reames13} and major geomagnetic storms as they arrive at Earth and interact with the terrestrial magnetosphere \citep[e.g.,][]{gosling90}.
Therefore, the capability of accurately describing and modeling solar eruptions
is of great importance from both theoretical and practical points of view. 
Specific attention has been paid to eruptions originating in active regions (ARs), as those produce the majority of CMEs \citep[][]{liu.l17}, as well as the fastest and, typically, the most geoeffective ones \citep[e.g.,][]{gopalswamy18}.

The energy required to power an eruption is gradually accumulated in the corona and stored in current-carrying, approximately force-free magnetic fields. These pre-eruptive configurations (PECs) are thought to be magnetic flux ropes (MFRs), sheared magnetic arcades, or some hybrid between the two \citep[e.g.,][]{patsourakos20}. The evolution of an eruption typically occurs in three consecutive phases, consisting of a slow initiation phase, an impulsive main phase, and a final propagation phase \citep[e.g.,][]{zhang06}.
 
For most of the time, this evolution proceeds slowly enough to be adequately modeled via numerical magnetohydrodynamic (MHD) simulations.
To initialize MHD simulations of solar eruptions,
constructing PECs and setting appropriate boundary conditions are required.
For the modeling of observed events, both of these procedures
have to be constrained by observational data, of which photospheric magnetic data undoubtedly play a central role \citep[e.g.,][]{torok18}.
Modeling a realistic PEC that is close to a force-free equilibrium (due to
the strong magnetic fields present in ARs)
and constrained by observed magnetic data is itself a nontrivial problem.
Many methods have been developed for constructing
PECs with different degrees of realism and complexity, for both idealized and real-event cases. 
These include, in particular, (1) evolutionary methods based on boundary-driving, either
magnetofrictional \citep{Cheung2012, Price2020} or via slow MHD flows \citep{Amari2000, Lionello2002, linkeretal2003, Bisi2010, Zuccarello2012, Mikic2013a}, (2) nonlinear force-free field (NLFFF) extrapolations \citep[e.g.,][]{Schrijver2008,canou10, jiang18}, (3) MFR insertion method \citep{vanBall2004, Su2011, Savcheva2012}, and (4) analytic MFR-embedding models \citep{lugaz11, manchester14, Titov2014, Titov2018, torok18, downs21, kang23, Sokolov2023}.

Due to the inherent nonlinearity of the problem, multiple iterations or evolutionary steps may be required to produce a PEC that satisfactorily matches the magnetic data, irrespective of the used method.
During such iterations, it is convenient to treat the current-carrying and potential parts of the modeled PEC separately, at least as far as their contributions to the normal (or radial) field component, $B_{r}$, at the boundary are concerned.
The main purpose of this article is to provide a general approach that allows one to perform such a 
separate treatment, by excluding the contribution of coronal currents to the photospheric distribution of $B_{r}$, so that the latter is associated only with the potential magnetic field in the corona.

This representation of a coronal magnetic field is useful, at least, for solving the following two problems.
The first is the modeling of PECs through the superposition of an MFR and a given potential magnetic field, where the MFR is constructed using our previously developed regularized Biot\textendash{}Savart laws
(\RBSL{s}; see V. S. Titov et al. \citeyear{Titov2018}, \citeyear{Titov2021}). 
Now, the potential field must be calculated only once, which can benefit PEC modeling methods that involve trial-and-error or iterative optimization. The decomposition can also help guide the placement of the currents in the PEC model (e.g. the MFR geometry). Using our \RBSL{} method for modeling observed PECs \citep[][]{Titov2018} as an example, we demonstrate that this approach greatly advances such procedures.

The second problem is the decomposition of a given photospheric magnetic field $\Bv$ into the following three parts:
(1) a potential field $\Bpot$ derived from a given $B_{r}$ and generated by subphotospheric currents that do not reach the solar surface,
(2) a toroidal field $\BvT$ derived from a given $J_{r}$, and (3) a poloidal field $\BvSt \equiv \Bv - \Bpot - \BvT$.
The fields $\BvT$ and $\BvSt$ are purely tangential and are generated by all coronal currents, subphotospheric closure currents, and corresponding auxiliary fictitious sources described below.

These two problems are closely related, as the solution of the second problem actually helps to solve the first one. 
This is because it allows one to identify the location of MFRs in projection to the boundary, particularly their footprints, as well as the presence and direction of unneutralized currents in MFRs by using only vector magnetic data, that is, {\it before modeling PECs}.
These capabilities can significantly increase the power and precision of the \RBSL{} method.

The idea of decomposing a surface magnetic (terrestrial) field into internal and external sources originated with 
Gauß \citep{Gauss1839}.
Recently, \citet{Schuck2022} applied this idea to the solar case by implementing, with the help of spectral methods, the so-called Carl’s (Gauß) Indirect Coronal Current
Imager (CICCI) both for planar and boundaries.
\citet{Welsch2022X} reformulated this approach in terms of Fourier transforms for a planar boundary.
The innovative approaches of these articles can provide important insights into the nature of coronal currents and inspired us to look at vector decomposition in more detail.
The toroidal field $\BvT$ that arises in our decomposition is identical to theirs.  However, their other two parts, the poloidal magnetic fields $\BPlt$ and $\BPgt$, are very different compared to ours.
These two fields are purely {\it potential}  in the infinitesimal proximity to the corresponding sides of the boundary.
At first sight, this is in contrast to
the fact that the photospheric field $\Bv - \BvT$, in general, has to be {\it nonpotential}.
One reason for the latter is the sharp drop (on the hydrostatic length scale $\sim 100\; \mathrm{km}$) of the plasma pressure near the surface in a magnetic field that is generally 
nonradial, which must be accommodated by a nonvanishing transverse current density.

This contradiction, however, is only apparent and can be resolved by  realizing that the tangential components of $\BPlt$ and $\BPgt$ must be continuous across the surface if the corresponding current densities below and above it, respectively, are continuous.
A jump in current density across the surface associated with $\BPlt$ or $\BPgt$ can only cause a jump in the radial derivatives of the corresponding tangential components, but not in their values.
Thus, the use of $\BPlt$ and $\BPgt$ in the Gaussian decomposition 
is consistent with the above-mentioned nonpotentiality of the field $\Bv - \BvT$.
It should also be emphasized that, by construction, all three parts of the field in our method refer to the same side of the surface and
are determined without using their volumetric properties nearby.
These subtle differences between the two methods must be kept in mind when
applying them to the analysis of vector magnetic data,
in particular, to the practical problem of the determination of MFR locations and parameters.

The theoretical foundation of our decomposition method is based on our generalization of the Biot\textendash{}Savart law described in Section \ref{s:mmBSl}.
The method of decomposition and its application to modeled and observed PECs are presented in Section \ref{s:dvm}.
We demonstrate that this method indeed allows one to identify the MFR footprints as well as the MFR shape in projection to the photospheric surface, which provides important constraints for comprehensive modeling of PECs.
The results obtained are summarized in Section \ref{s:sum}.
Appendix \ref{s:dBC} thoroughly describes the derivation of the magnetic field of auxiliary fictitious magnetic sources and their physical meaning.
In Appendix \ref{s:relation}, we discuss in more detail the relationship between the two decomposition methods considered here.

\section{Magnetogram-matching Biot\textendash{}Savart Law
	\label{s:mmBSl}
}

\subsection{Preface
\label{ss:preface}}

In MHD, the magnetic field and electric current are linearly coupled via Ampere's law, whose differential form states that the current density is the curl of the magnetic field.
For our purpose, however, we need to use the inverted version of this relationship, provided by Biot\textendash{}Savart's law (\BSL).
By applying the latter to a given volumetric distribution of the current density, one can determine the corresponding magnetic field.
Note that the normal component of this field will generally not vanish at the boundary.
However, \citet{Isenberg2007} noticed that, for the case of a coronal line current, this component can be reduced to zero by adding a subphotospheric current that closes the current path. We refer to this subphotospheric current as ``closure current'' henceforth.
They achieved this by approximating the boundary surface with a plane and using the mirror image of the coronal-current path about this plane as the closure-current path.

A similar approach in planar geometry, but for distributed coronal currents, was used earlier by \citet{Wheatland2004, Wheatland2007} in his numerical method for NLFFF extrapolations of the surface magnetic field to the corona.
He presented a Fourier solution for the current-carrying part of the magnetic field in Cartesian coordinates via imposing $B_z=0$ as a boundary condition, rather than explicitly constructing it by using a mirror current.
\citet{Gilchrist2014} later generalized this method to spherical geometry, using a global representation of the magnetic field with vector spherical harmonics.
By extending the concept of mirror currents, our article provides a similar generalization for spherical geometry for both line and distributed currents, the closure-current paths of which can be of arbitrary shape.
These extensions should be particularly useful for modeling PECs that contain elongated filament channels.

Note that mirroring a coronal current about the surface boundary is not just a mathematical ``trick,'' but rather a natural way to incorporate the consequences of photospheric MHD line-tying conditions directly into \BSL.
These conditions take into account the effect of density stratification on ideal MHD perturbations of magnetic fields \citep[e.g.,][]{Hood1986}.
With respect to such perturbations caused, e.g., by solar eruptions, they strongly idealize the inertia of the dense plasma at and below the photosphere by treating this region of the Sun as an ideal rigid conductor.

Following this idealized approach, we henceforth assume that the dense solar interior is separated from the tenuous corona by the photospheric surface whose elements possess an infinite inertia, so they cannot be set in motion by coronal MHD flows.
Due to the frozen-in-law condition, these flows can then change only the tangential, but not the radial, component of the magnetic field at the photospheric boundary.

Note, in this respect, that solar eruptions involve relatively fast MHD flows above the surface, so a significant build-up of the tangential magnetic field is expected in its vicinity.
The latter, in turn, implies the cumulation of the current density tangential to the surface and generally transverse to the magnetic field.
In a more realistic approach that takes into account a finite inertia of the interior plasma, such a cumulation of the current cannot be persistent.
This is because the photospheric cross-field current is likely to propagate via torsional Alfv\'en waves downward to the convection zone \citep{Melrose1995}.
Thus, the line-tying effect can act only if this propagation lasts longer than the eruption.
Bearing in mind that the interior plasma is $\sim\!\! 4$ orders of magnitude denser than the coronal plasma, we will further assume that this condition on the respective time scales is satisfied at least for the dynamic phase of eruptions.

Motivated by these considerations, we derive here a new form of \BSL{} that allows one,
for a given line or distributed closed current, to calculate the coronal magnetic field with a vanishing $B_{r}$ at the spherical boundary.
Thus, if one superimposes such a modified \BSL{} field with the potential magnetic field determined from, say, an observed boundary distribution of $B_{r}$, the total current-carrying magnetic field will match this distribution.
Therefore, we now use the term {\it  magnetogram-matching \BSL} (\MBSL).

We achieve this by including into the classical \BSL{}  elementary potential fields produced by auxiliary fictitious sources, all located within the solar interior.
The sources are represented by magnetized shells of triangular shape whose one vertex is situated at the center of the Sun and the other two below the boundary, at an infinitesimal distance from each other.
By construction, such a shell
produces a potential field whose $B_{r}$ at the boundary compensates that of a \BSL{} current element, regardless of where this element is located---below or above the boundary.
We exploit here essentially the same approach as the one used in the method of images for solving magnetostatic and electrostatic problems \citep[e.g.,][]{Jackson1962}.

Applying our \MBSL{} approach only to the subphotospheric currents that provide a closure to coronal current loops, which are either a single-line current or a continuum of current tubes with infinitesimal cross sections, we show that their total field, both in the coronal volume and at the surface, depends only on the foot points of these loops.
Moreover, it is actually part of the so-called toroidal magnetic field, which has a vanishing $B_{r}$ and, summed with the poloidal field, forms the entire coronal configuration (see \citealt{Schuck2022} and \citealt{Yi2022} and references therein).
We substantiate this nontrivial result by means of two complementary proofs: one is purely mathematical, while the other relies on the physical properties of the fictitious sources that produce the compensating magnetic field for the subphotospheric current elements. 

In contrast, the application of \MBSL{} to coronal currents provides a magnetic field whose $B_{r}$ generally does not vanish in volume and depends on the shapes of the current paths.
By adding it to the toroidal field and the potential field derived from the photospheric $B_{r}$, one obtains the entire coronal configuration.
The current distribution associated with the toroidal field here provides the required closure for the coronal-current loops in this configuration.
It matches the radial component of the current density, $J_{r}$, at the boundary without affecting $B_{r}$ there.

In contrast to \citet{Schuck2022}, whose field $\BPgt$ has a nonvanishing $B_{r}$ at the surface, our decomposition method allows
the photospheric field associated with the coronal currents to be purely tangential to the boundary.
Such a field, taken alone, would be produced on the surface in response to the induction of the coronal currents if the solar globe were an ideal rigid conductor.
The photospheric distribution of $B_{r}$ can then be associated only with currents that circulate solely in the interior.
In light of the above discussion, our type of decomposition, compared to the Gaussian one, appears to be more appropriate for analyzing transitions from pre-eruptive to post-eruptive magnetic configurations.

Thus, the application of \MBSL{} to both the interior and the corona of the Sun allows one to separate the current-carrying part of the configuration such as if its coronal \BSL{} field were completely shielded
at the boundary by surface currents.
The latter could even be determined from the resulting tangential field component by assuming that this component becomes zero when crossing the solar surface toward the interior.
Using these surface currents instead of our fictitious sources would provide another way to derive the \MBSL.
However, as already mentioned, they are not a purely abstract construct, as in our field decomposition.
Rather, a part of these surface currents have to develop during eruptions, because of the line-tying effect, in response to relatively fast variations of volumetric coronal currents.
This is consistent with both existing MHD models and observations, which show much larger changes in the photospheric transverse field $\Bv_{\mathrm{t}}$ than in $B_{r}$ during eruptions \citep[e.g.,][]{wang1992, wangetal1994,sun17}.

\subsection{Closed-line Current
	\label{ss1:mmBSl}
}

\begin{figure}[ht!]
\centering
\resizebox{0.55\textwidth}{!}{
\includegraphics{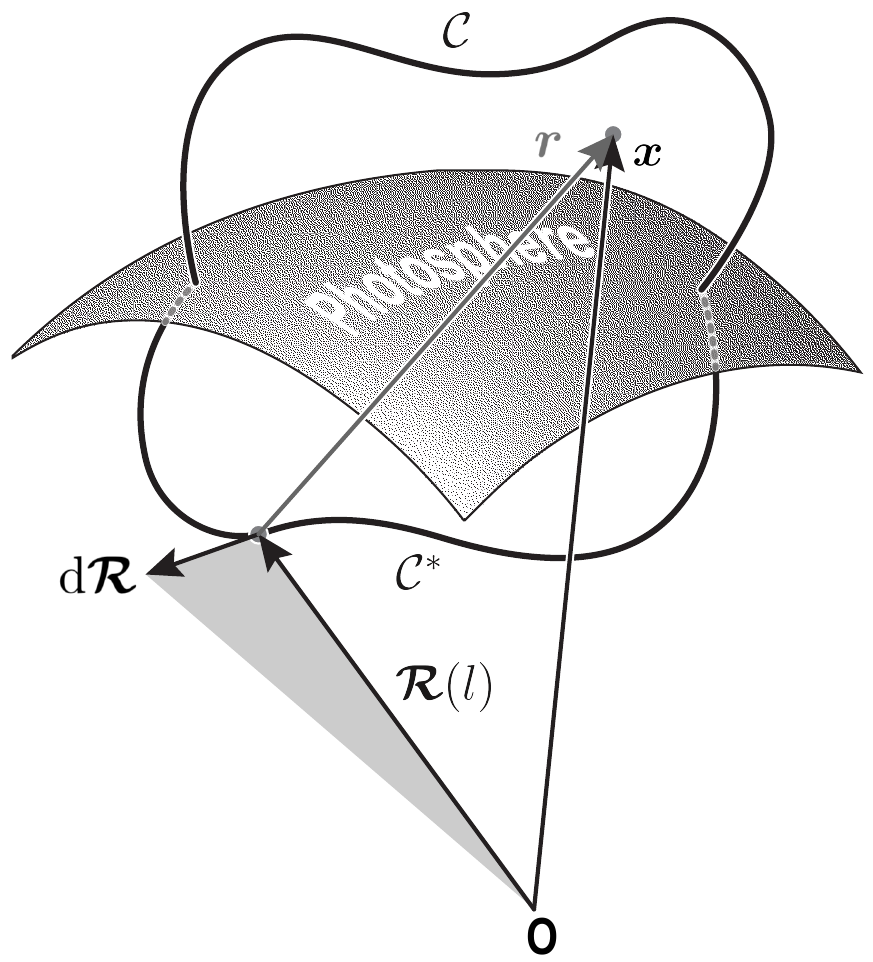}}
\caption{
The radius vector $\Rv(l)$ with the starting point at the center of the Sun {\bf O} defines a line-current path that consists of coronal and subphotospheric parts, $\Cc$ and $\Cs$, respectively.
The integration of the elementary Biot\textendash{}Savart field (see Equation (\ref{dBI})) along this path provides the potential magnetic field produced by the current circuit at a given point $\xv$.
	\label{f:path}
}
\end{figure}

Consider a line current of strength $I$ flowing along a closed path $\Cc \cup \Cs$, where $\Cc$ and $\Cs$ are its parts above and below, respectively, the solar surface of radius $R_\sun$ (Figure \ref{f:path}).
Let this path be represented by the radius vector $\Rv(l)$ and parameterized by the arc length $l$ measured from one of the foot points, so that $\Rvp = {\mathrm d}\Rv/{\mathrm d}l$ is a unit vector tangential to the path.
Then, according to \BSL, the infinitesimal contribution of a path element of length ${\mathrm d}l$ to the magnetic field $\bm{B}_{I}$ produced by $I$ at a given observation point $\xv$ is described by
\begin{eqnarray}
\left[
\frac
{\mu I}
{4\pi R_{\sun}}
\right]
\qquad
{\mathrm d}
\bm{B}_{I}
&=&
-
\frac
{
\left(
\xv
-
\Rv
\right)
\bmt
\Rvp
}
{
\left|
\xv
-
\Rv
\right|
^3
}
{\mathrm d}l
=
\frac
{
\rv
\bmt
{\mathrm d}
\rv
}
{
r^3
}
=
\frac
{
\hat{\rv}
\bmt
{\mathrm d}
\rv
}
{
r^2
}
\,
,
	\label{dBI}
\\
\rv
&=&
\xv
-
\Rv
\,
,
\quad
\hat{\rv}
=
\rv
/
r
\,
,
	\label{rv}
\end{eqnarray}
where the expression in the brackets on the left-hand side of Equation (\ref{dBI}) represents the unit in which $\bm{B}_{I}$ is measured.
Similar expressions in the brackets will be used further for designating the units of other values in our paper.
Also, we assume hereafter that $l$ and the lengths of all vectors are normalized to $R_{\sun}$ and $\mu$ is the magnetic permeability in vacuum.

The result of integration of ${\mathrm d} \bm{B}_{I}$ along the path is a potential magnetic field that for $r> a/R_{\sun}$ approximately describes the field of an MFR with a circular cross section of radius $a$ \citep{Titov2018}.
This field generally has a nonvanishing radial component at the photospheric boundary.
If the characteristic size of the configuration is smaller than $R_{\sun}$, then the value of this component can be significantly reduced by defining the subphotospheric path $\Cs$ as a mirror image of $\Cc$ about the plane that locally approximates the boundary surface \citep[][]{Titov2021}.
However, the deviation of the radial component from zero at the spherical boundary is growing with size, and so it is desirable to make it disappear.
Otherwise, the field superposed of $\bm{B}_{I}$ and the ambient potential field would not match the corresponding magnetogram.

\subsubsection{ The Inversion and Closure of the Path $\Cc$
	\label{sss:inversion} }

In spherical geometry, the analog of mirroring a point $\Rv$ about the boundary is the inversion transformation \citep[see, e.g,][]{Coxeter1969} given by
\begin{eqnarray}
\Rv_{*}
=
\Rv
/
{\mathcal R}^2
\,
,
	\label{Rvs}
\end{eqnarray}
where $\Rv_{*}$ is the image of $\Rv$.
Using this formula and the fact that the direction of the current in the image should be switched to the opposite, one can obtain the following relationship between the radial components of Equation (\ref{dBI}) and its transformed expression ${\mathrm d}\bm{B}^{*}_{I}$ at the boundary: 
\begin{eqnarray}
\Rvu
\bcd
\left.
{\mathrm d}\bm{B}^{*}_{I}
\right|
_{\xv=\Rvu}
=
-
{\mathcal R}_{*}
\,
\Rvu
\bcd
\left.
{\mathrm d}\bm{B}_{I}
\right|
_{\xv=\Rvu}\,
,
	\label{RdBIKt}
\end{eqnarray}
where
$
{\mathcal R}_{*}
\equiv
\left|
\Rv_{*}
\right|
$
and $\Rvu$ is the radius vector of a given point at the photospheric boundary such that $|\Rvu|=1$, as normalized to $R_\sun$.
Equation (\ref{RdBIKt}) makes it clear that these components cancel each other only in the limit of ${\mathcal R} \rightarrow 1$.
Thus, the trick with mirroring the path $\Cc$ about the boundary works for spherical geometry only approximately, even if the mirroring is made via the inversion.

In principle, the problem under consideration can be solved with better accuracy by suitably modifying the observed magnetogram as a boundary condition for the calculated potential field, so that the sum of this field with $\bm{B}_{I}$ would match the original magnetogram \citep{Titov2018}.
However, the exact shape of the modeled MFR is a priori unknown and therefore finding a sufficiently accurate solution may require multiple, numerically expensive iterations.
In the following, we propose an alternative method for solving this problem exactly, which is based on a new form of the \BSL, all elementary magnetic fields of which are strictly tangential to the spherical boundary.

\subsubsection{Nonpotentiality of the Elementary \BSL{} Field 
	\label{sss:eBSLnonpot}}

Before describing our new method, note first that the elementary fields ${\mathrm d}\bm{B}_{I}$ given by Equation (\ref{dBI}) are not potential.
Indeed, the curl of this equation yields the following result:
\begin{eqnarray}
\bm{\nabla}_{\xv}
\bmt
{\mathrm d}
\bm{B}_{I}
=
\frac
{
{\mathrm d}
\rv
}
{
r^3
}
-
3
\frac
{
\left(
\rv
\bcd
{\mathrm d}
\rv
\right)
}
{
r^5
}
\rv
=
\left(
{\mathrm d}
\rv
\!
\bcd
\!
\bm{\nabla}_{\rv}
\right)
\!
\frac
{\rv}
{r^3}
,
	\label{cdBI}
\end{eqnarray}
which shows that, by necessity of the charge conservation, each individual infinitesimal current element is formally closed via nonvanishing volumetric currents.
The latter are as imaginary as an isolated current element and distributed in space as a potential field of the dipole
$
{\mathrm d}
\rv
=
-
{\mathrm d}
\Rv
$.
Since Equation (\ref{cdBI}) is the total differential of the vector $\rv/r^3$, its integration over the entire closed path yields a vanishing current density in the volume.
This is a result of the mutual cancellation of the current dipoles having a ``head-to-tail'' distribution along the path.
Due to this cancellation, the corresponding integration of ${\mathrm d}\bm{B}_{I}$ over the same closed path provides a purely potential field $\bm{B}_{I}$ outside.
In contrast, the field obtained by the integration of ${\mathrm d}\bm{B}_{I}$ over a part of this path is never potential.

This fact is valid, of course, for distributed current systems as well, since they can be represented as a continuum of current tubes with infinitesimally thin cross sections, the \BSL{} contributions of which to the total magnetic field are linearly superimposed.
Therefore, one should pay attention to the connectivity of such tubes with respect to the photospheric boundary in order to figure out whether their contributions in the region of interest are curl-free or not.

\subsubsection{ Compensating Potential Magnetic Fields 
	\label{sss:dBCdBCs}}

It turns out that, for every elementary field ${\mathrm d}\bm{B}_{I}$, there exists a corresponding potential magnetic field whose sources are located at $|\xv|\le1$ and whose radial component at the boundary equals
$
-
\Rvu
\bcd
\left.
{\mathrm d}\bm{B}_{I}
\right|
_{\xv=\Rvu}
$.
We call this potential field the compensating field and denote it as
$
{\mathrm d}
\bm{B}
_{\Cs}
$
or
$
{\mathrm d}
\bm{B}
_{\Cc}
$
depending on whether the corresponding current element belongs to the path $\Cs$ or $\Cc$, respectively.
As described at length in Appendix \ref{s:dBC}, these fields have the following expressions:
\begin{eqnarray}
&&
\left[
\frac
{\mu I}
{4\pi R_{\sun}}
\right]
\qquad
{\mathrm d}
\bm{B}
_{\Cs}
=
-
\frac
{
\hat{\rv}
\bmt
{\mathrm d}
\rv
}
{
r^2
}
+
\frac{
\left( 
\hat{\rv}
\bcd
{\mathrm d}
\rv
\right)
\,
\hat{\rv}
\bmt
\hat{\xv}
}
{
r^2
\,
\left(
\hat{\rv}
\bcd
\hat{\xv}
+
1 
\right)
}
\nonumber
\\
&&
\qquad
\qquad
\qquad
+
\left(
\frac
{1}
{r}
+
\frac
{1}
{|\xv|}
\right)
\,
\frac{ 
\left[
\left(
\hat{\rv} 
+
\hat{\xv}
\right)
\bmt
\left(
\hat{\rv}
\bmt
{\mathrm d}\rv
\right)
\right]
\bmt
\hat{\xv}
}
{
r
\,
\left(
\hat{\rv}
\bcd
\hat{\xv}
+
1 
\right)^2
}
\,
,
	\label{dBCs}
\end{eqnarray}
\begin{eqnarray}
&&
\left[
\frac
{\mu I}
{4\pi R_{\sun}}
\right]
\qquad
{\mathrm d}
\bm{B}
_{\Cc}
=
-
\frac
{
\hat{\rv}_{*}
\bmt
{\mathrm d}
\rv
}
{
{\mathcal R}
r_*^2
}
+
\frac{
\left( 
\hat{\rv}_{*}
\bcd
{\mathrm d}
\rv
\right)
\,
\hat{\rv}_{*}
\bmt
\hat{\xv}
}
{
{\mathcal R}
\,
r_*^2
\,
\left(
\hat{\rv}_{*}
\bcd
\hat{\xv}
+
1 
\right)
}
\nonumber \\
&&
\qquad
\qquad
\qquad
+
\left(
\frac
{1}
{r_*}
+
\frac
{1}
{|\xv|}
\right)
\,
\frac{ 
\left[
\left(
\hat{\rv}_{*}
+
\hat{\xv}
\right)
\bmt
\left(
\hat{\rv}_{*}
\bmt
{\mathrm d}\rv
\right)
\right]
\bmt
\hat{\xv}
}
{
{\mathcal R}
\,
r_*
\,
\left(
\hat{\rv}_{*}
\bcd
\hat{\xv}
+
1 
\right)
^2
}
\,
,
	\label{dBC}
\end{eqnarray}
where
\begin{eqnarray}
\rvs
&=&
\xv
-
\Rvs
\,
,
\quad
\hat{\rv}_{*}
=
\rvs
/
r_{*}
\,
,
\\
\hat{\xv}
&=&
\xv
/
|\xv|
\,
.
	\label{rvs}
\end{eqnarray}

\begin{figure*}[ht!]
\centering
\resizebox{0.99\textwidth}{!}{
\includegraphics{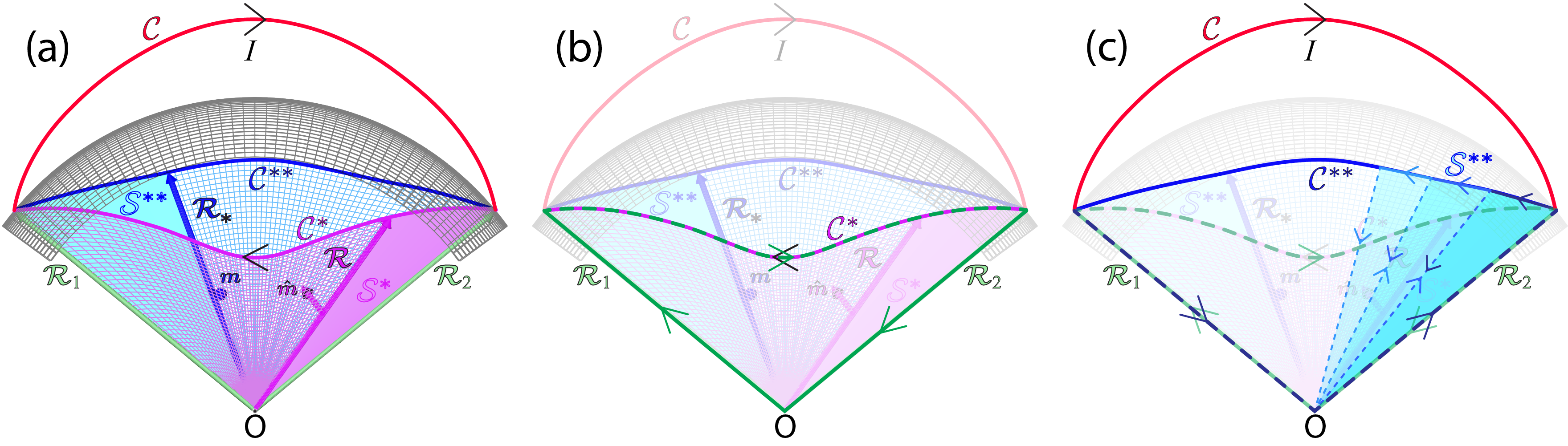}}
\caption{
(a) The construction of the compensating potential magnetic field for each of the two line-current paths, one of which, $\Cs$ (magenta), is situated below the photospheric boundary (gray grid) and the other, $\Cc$ (red), above it.
For the path $\Cs$ ($\Cc$),  this field is generated by a magnetic shell whose magnetic moment $\mvh$ (magenta)  ($\mv$ (blue)) is distributed over and perpendicularly to a ruled surface $\Ss$ (magenta) ($\Sss$ (cyan)).
This ruled surface is swept out by the radius vector $\Rv$ ($\Rv_{*}$) as its head slides along the path $\Cs$ (magenta) (the curve $\Css$ (blue)) from the foot point $\Rv_{2}$ to $\Rv_{1}$ (vice versa).
The curve $\Css$ is the image of the path $\Cc$ produced by the inversion mapping (see Equation (\ref{Rvs})).
(b) Due to a uniform distribution of $\mvh$ in $\Ss$, the superposition of the corresponding elementary surface currents results in the line current flowing along the edge of $\Ss$ (solid green and dashed lines).
As this edge line current and the original current in the entire path $\Cc \cup \Cs$ have the same value and circulation, they cancel each other out on the path $\Cs$.
The remaining edge current at the straight edges of $\Ss$ generates the toroidal magnetic field $\Bv_{I\Cs}$ given by Equation (\ref{dBICs}).
(c) Due to a nonuniform, proportional to ${\mathcal R}_{*}^{-1}$,
distribution of $\mv$ in $\Sss$, the corresponding surface
current density does not vanish in $\Sss$ to maintain the appropriate variation of the edge line current along the curve $\Css$.
The variation of $|\mv|$ and associated edge currents in three infinitesimal triangles of $\Sss$ is represented by varying hues of cyan and blue, respectively.
	\label{f:ScSs}}
\end{figure*}

Note that only the first terms of Equations (\ref{dBCs}) and (\ref{dBC}) have nonvanishing radial components, because their other terms have forms of the vector product with $\hat{\xv}$.
Taking then the scalar product of $\hat{\xv}$ with Equation (\ref{dBCs}) and using Equation (\ref{dBI}), one immediately obtains that
\begin{eqnarray}
\hat{\xv}
\bcd
{\mathrm d}
\bm{B}_{\Cs}
=
-
\hat{\xv}
\bcd
{\mathrm d}\bm{B}_{I}
\,
.
	\label{dBICs*xv=0}
\end{eqnarray}
This means that  ${\mathrm d} \bm{B}_{\Cs}$ fully compensates for the radial component of ${\mathrm d} \bm{B}_{I}$ generated by a subphotospheric current element not only at the boundary but also everywhere in the coronal volume.

Similarly, the scalar product of $\Rvu=\hat{\xv}$ and the first term of Equation (\ref{dBC}), after restricting it with the help of Equation (\ref{Rvs}) for the boundary surface, yields
\begin{eqnarray}
\Rvu
\bcd
{\mathrm d}
\Bv_{\Cc}
\est{\xv=\Rvu}
=
-
\Rvu
\bcd
{\mathrm d}\bm{B}_{I}
\est{\xv=\Rvu}
\,
.
	\label{bcC}
\end{eqnarray}
This means that ${\mathrm d} \bm{B}_{\Cc}$ fully compensates the photospheric radial component of ${\mathrm d} \bm{B}_{I}$ generated by a coronal-current element, as required.

The compensating magnetic field ${\mathrm d} \Bv_{\Cs}$ admits a simple physical interpretation derived in Appendix \ref{A1}.
This implies that this field is generated by infinitesimal magnetized triangles spanned on vectors $\Rv$ and their head displacements
$
{\mathrm d}
\Rv
=
-
{\mathrm d}
\rv
$
along the path $\Cs$ (see Figure \ref{f:path}).
Every such triangle has a magnetic moment that is uniformly distributed with a surface density $\mvh$ over its area and perpendicular to its plane to form an elementary magnetized, or magnetic, shell  \citep[see textbooks; e.g.,][]{Stratton1941} .

The total compensating field for the path $\Cs$ is therefore produced by a magnetic shell that is assembled from all such magnetized triangles that abut the path $\Cs$.
In other words, the assembled magnetic shell is a uniformly and orthonormally magnetized ruled surface $\Ss$ whose directrix is the path $\Cs$.
This surface is swept out by the radius vector $\Rv$ as its head moves along $\Cs$ from the foot point $\Rv_{2}$ to $\Rv_{1}$.
The result is a curvilinear triangle $\Ss$ that has two straight sides and one curved side (see Figure \ref{f:ScSs}(a)).
Due to a constant value of $|\mvh|$, the corresponding elementary surface currents mutually cancel each other throughout $\Ss$, except for the edges of $\Ss$, where they superpose into a nonvanishing line current \citep{Stratton1941}.

The compensating magnetic field ${\mathrm d} \Bv_{\Cc}$ has more sophisticated underlying physics, which Appendix \ref{A2} describes in detail.
This field is also generated by a magnetic shell that is positioned, however, on a different ruled surface $\Sss$.
The directrix of $\Sss$ is a curve $\Css$ obtained from $\Cc$ with the help of the inversion mapping whose point-wise definition is given by Equation (\ref{Rvs}).
As in the case of $\Ss$, the magnetic moment surface density field $\mv$ is orthogonal to $\Sss$.
However, its modulus $|\mv|$ is constant only along the dimension of $\Sss$ that is parallel to the vectors $\Rvs$.
Along its second, transversal to $\Rvs$, dimension, $|\mv|$ varies proportionally to ${\mathcal R}^{-1}_{*}$, which leads to the presence of a nonvanishing surface current density in $\Sss$.

The described physical meaning of the compensating magnetic fields will help us thoroughly understand the properties of the field produced by a coronal line current under the condition that its radial component vanishes at the photospheric boundary.

\subsubsection{ Toroidal Magnetic Field and Flux Function 
	\label{sss:Btor}}

It is interesting that the numerical integration of the field
\begin{eqnarray}
{\mathrm d} \bm{B}_{I\Cs}
\equiv
{\mathrm d} \bm{B}_{I}
+
{\mathrm d} \bm{B}_{\Cs}
	\label{dBICs}
\end{eqnarray}
demonstrates that the integrated field $\bm{B}_{I\Cs}$ practically does not depend on the shape of the path $\Cs$!
This surprising result motivated us to search for its mathematical proof, which was established by discovering that Equation (\ref{dBICs}) considered at fixed $\xv$ as a function of $\rv$ is actually represented by the following total differential:
\begin{eqnarray}
{\mathrm d}
\bm{B}_{I\Cs}
&=&
-
\left(
{\mathrm d}\rv
\bcd
\bm{\nabla}_{\rv}
\right)
\bm{\Theta}(\xv,\rv)
\, ,
	\label{dBICs2}
\\
\bm{\Theta}
(
\xv
,
\rv
)
&=&
\left(
\frac
{1}
{r}
+
\frac
{1}
{|\xv|}
\right)
\,
\frac{ 
\hat{\rv}
\bmt
\hat{\xv}
}
{
\hat{\rv}
\bcd
\hat{\xv}
+
1
}
\,
.
	\label{Thev}
\end{eqnarray}
In other words, $\bm{\Theta}(\xv,\rv)$ is an indefinite vector-valued integral of the vector field defined by Equation (\ref{rv}) where the variable $\Rv$ represents all possible subphotospheric paths $\Cs$ and $\xv$ is a fixed point given such that $|\xv|\ge 1$.
Thus, the integration of ${\mathrm d}\bm{B}_{I\Cs}$ along any of these paths yields the corresponding definite integral
\begin{eqnarray}
\Bv_{I\Cs}
&=&
{
\bm{\Theta}(\xv,\rv)
\bigr|}
_{\rv=\xv-\Rv_1}
^{\rv=\xv-\Rv_2}
	\label{BICs}
\, ,
\end{eqnarray}
where $\Rv_1$ and $\Rv_2$ are the second (first)  and first (second) foot points, respectively, of the path $\Cs$ ($\Cc$).
In accordance with our numerical examples, the integrated field $\Bv_{I\Cs}$, in fact,
depends on the foot points of the path $\Cs$, but not on its shape.

This nontrivial result can easily be understood if one turns to the physics behind the compensating magnetic field for the path $\Cs$.
As shown in Figure \ref{f:ScSs}(a) and described in Section \ref{sss:dBCdBCs}, this field is generated by a uniformly magnetized shell $\Ss$ whose surface currents superpose into a line current circulating along the sides of $\Ss$.
Appendix \ref{A1} shows that the value and direction of circulation of this current are the same as for the original current flowing along the path $\Cc \cup \Cs$.
It is obvious then that these currents cancel each other on the path $\Cs$ (see Figure \ref{f:ScSs}(b)).
Thus, the field $\Bv_{I\Cs}$ is generated only by the current flowing along the straight sides of $\Ss$ and therefore does not depend on the shape of $\Cs$.
We checked that the integration of the elementary \BSL{} field given by Equation (\ref{dBI}) yields indeed the same Equations (\ref{Thev}) and (\ref{BICs}).

The field $\bm{B}_{I\Cs}$ is not potential because it is not curl-free.
However, the calculations show that the curl of this field is a potential field given by the following gradient:
\begin{eqnarray}
\bm{\nabla}_{\xv}
\bmt
\bm{B}_{I\Cs}
&=&
\bm{\nabla}_{\xv}
\left(
\frac
{1}
{|\xv-\Rv_2|}
-
\frac
{1}
{|\xv-\Rv_1|}
\right)
\,
.
\nonumber
\\
	\label{rotBICs}
\end{eqnarray}
Thus, $\Bv_{I\Cs}$ can equivalently be interpreted as the magnetic field generated by distributed coronal currents
 under the constraint $\Rvu \bcd \bm{B}_{I\Cs}\bigr|_{\xv=\Rvu} = 0$.
The distribution of these currents is defined by the gradient of a potential that is produced by
positive and negative point sources located at foot points of the path $\Cs$.
This result well matches the fact that the field $\Bv_{I\Cs}$ is generated by the edge current on the straight sides of $\Ss$.
The distributed current defined by Equation (\ref{rotBICs}) simply provides a closure for this edge current.

Taking into account that $\Bv_{I\Cs}$ has no radial component (see Equations (\ref{dBICs*xv=0}) and (\ref{Thev})), we managed to uncurl Equation (\ref{BICs}) and derive the corresponding vector potential
\begin{eqnarray}
\bm{A}_{I\Cs}
=
\hat{\xv}
\,
\mathcal{T}(\xv,\rv)
\bigr|
_{\rv=\xv-\Rv_1}
^{\rv=\xv-\Rv_2}
	\label{AICs}
\,
,
\end{eqnarray}
where
\begin{eqnarray}
\mathcal{T}(\xv,\rv)
=
\ln
\left(
\xv
\bcd
\rv
+
|\xv|
r
\right)
\,
.
	\label{Txr}
\end{eqnarray}
Taking the curl of Equation (\ref{AICs}), one can verify that
\begin{eqnarray}
\bm{B}_{I\Cs}
&=&
\bm{\nabla}
\bmt
\bm{A}_{I\Cs}
	\label{BICs2}
\, ,
\end{eqnarray}
as required.
Equations (\ref{AICs}) and (\ref{BICs2}) explicitly state that $\bm{B}_{I\Cs}$ is a toroidal magnetic field \citep{Schuck2022},
which simply means that $\bm{B}_{I\Cs}$ is solenoidal and tangential to spherical surfaces $|\xv|=\mathrm{const}$.
\begin{figure}[ht!]
\centering
\resizebox{0.55\textwidth}{!}{
\includegraphics{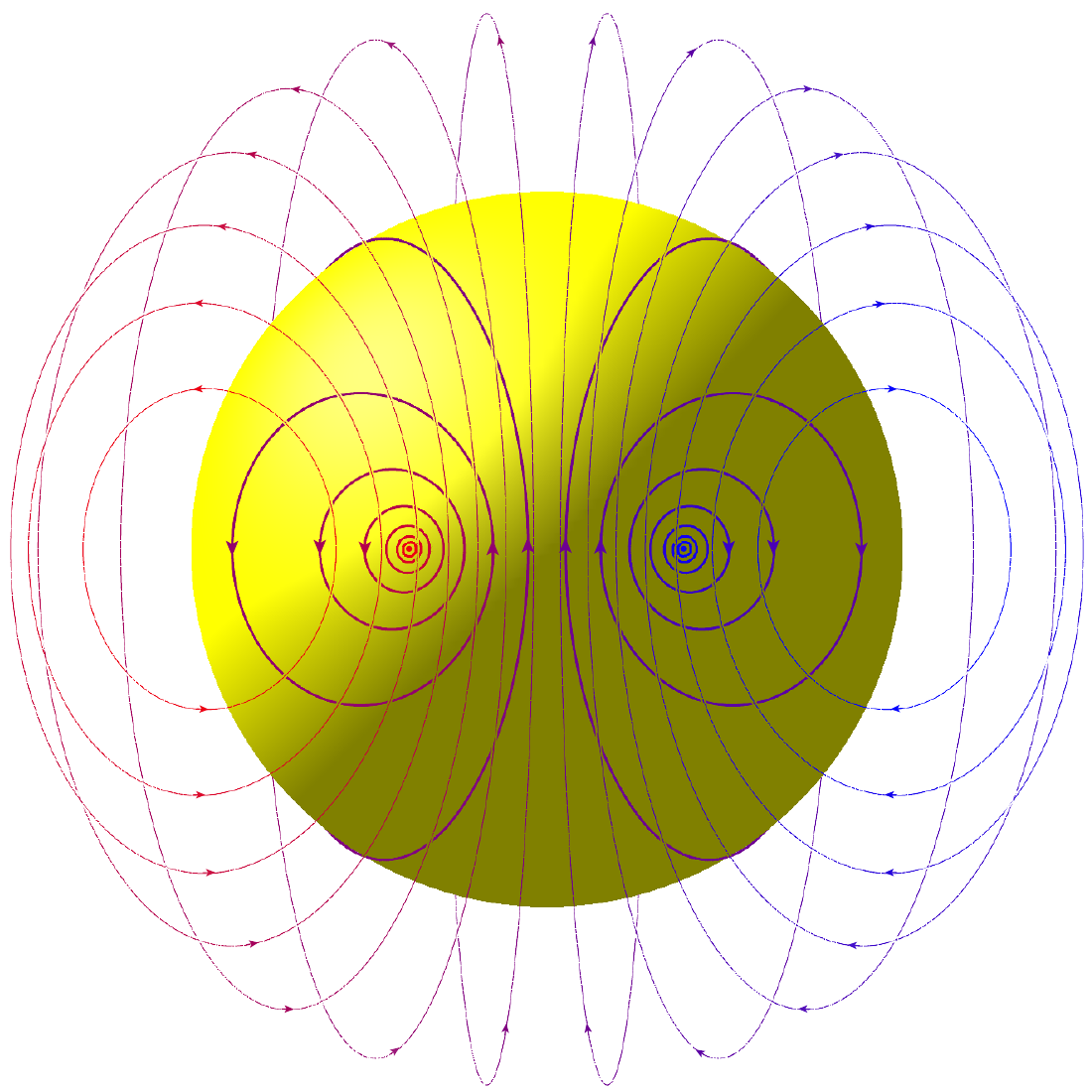}}
\caption{
Iso-contours of the TFFF given by Equation (\ref{f}) on two different surfaces: $|\xv|=1$ (thick lines on the yellow sphere) and $|\xv|=1.5$ (thin lines); the foot points of the corresponding current path are separated by minor-arc angle equal to $\pi/4$.
	\label{f:tff_conts}
}
\end{figure}

The scalar field
$
f(\xv,\Rv_1,\Rv_2)
\equiv
\mathcal{T}(\xv,\rv)
\bigr|
_{\rv=\xv-\Rv_1}
^{\rv=\xv-\Rv_2}
$
restricted on such a surface plays the role of the flux function for $\bm{B}_{I\Cs}$, because the magnetic flux of this field through a line element $\mathrm{d} \bm{l}$ tangential to the surface is
$
\bm{B}_{I\Cs}
\bcd
\mathrm{d}
\bm{l}
\bmt
\hat{\xv}
=
\mathrm{d}
\bm{l}
\bcd
\bm{\nabla}
f
$.
Using Equation (\ref{Txr}) and the foot point constraint,
$\Rv_1^2=\Rv_2^2=1$, one obtains
\begin{eqnarray}
&&
f
(
\xv
,
\Rv_1
,
\Rv_2
)
\bigr|
_{|\xv|=\mathrm{const}}
=
\ln
\left[
1
-
\frac
{
\hat{\xv}
\bcd
\Rv
}
{
|\xv|
}
+
\left.
\left(
1
-
2
\frac
{
\hat{\xv}
\bcd
\Rv
}
{
|\xv|
}
+
\frac
{
1
}
{
|\xv|^2
}
\right)
^{1/2}
\right]
\right|
^{\Rv=\Rv_2}
_{\Rv=\Rv_1}
\,
.
	\label{f}
\end{eqnarray}

The iso-contours of this toroidal field flux function (TFFF) represent magnetic field lines of $\bm{B}_{I\Cs}$, which all are nested in spherical surfaces $|\xv|=\mathrm{const}$ (see Figure \ref{f:tff_conts}).

\subsubsection{ Magnetogram-matching \BSL{} 
	\label{sss:mmBSL}}

Let us combine our results, given by Equations (\ref{dBI}), (\ref{dBC}), (\ref{Thev}), and (\ref{BICs}), to express the full coronal magnetic field $\bm{B}_{I\ominus}$ as follows:
\begin{eqnarray}
\bm{B}_{I\ominus}
=
\int
_{\Cc}
\left(
{\mathrm d}
\bm{B}
_{I}
+
{\mathrm d}
\bm{B}
_{\Cc}
\right)
+
\bm{\Theta}
(\xv,\rv)
\bigr|
_{\rv=\xv-\Rv_1}
^{\rv=\xv-\Rv_2}
\,
.
\nonumber
\\
	\label{mmBSL}
\end{eqnarray}
As shown above, this field is generated by the line current $I$ flowing along a closed path $\Cc \cup \Cs$ (Figure \ref{f:path}) under the condition that its radial component vanishes at the photospheric boundary.
$\bm{B}_{I\ominus}$ is normalized to $\mu I/(4\pi R_\sun)$ and therefore depends on the current $I$, radius $R_\sun$, and shape of $\Cc$, but not on the shape of $\Cs$, whose role is reduced to providing only the closure for the coronal current.

If we superpose $\bm{B}_{I\ominus}$ on a given ambient potential field, the radial component of the composite field will remain unchanged at the boundary. 
Therefore, Equation (\ref{mmBSL}) indeed represents the \MBSL{} for determining the field of the current-carrying path $\Cc$ that resides in such composite configurations.
For the paths tracking the MFR shapes, these configurations approximate the magnetic field outside current-unneutralized thin MFRs, which constitutes the major part of the coronal volume.

\subsubsection{ Application to Modeling \RBSL{} MFRs 
	\label{sss:mmBSL1rt}}

It is of interest for modeling \RBSL{} MFRs to derive another, slightly reduced form of the \MBSL.
As shown in Figure \ref{f:ScSs} and Appendix \ref{A3}, Equation (\ref{mmBSL}) admits a significant reduction: A part of its first term cancels its second term, which means that a part of the compensating field produced by the line current at the straight edges of the magnetic shell $\Sss$ eliminates the toroidal magnetic field.
The explicit expression for such a reduced form of the \MBSL{} is written as follows:
\begin{eqnarray}
&&
\left[
\frac
{\mu I}
{4\pi R_{\sun}}
\right]
\qquad
\bm{B}_{I\ominus}
=
\int_{\Cc} 
\frac
{
\hat{\rv}
\bmt
{\mathrm d}
\rv
}
{
r^2
}
\nonumber
\\
&&
+
\int_{\Css}
\left\{
\frac
{1}
{\mathcal R}
\frac
{
\hat{\rv}
\bmt
{\mathrm d}
\rv
}
{
r^2
}
\right.
+
\left(
1
-
\frac
{1}
{\mathcal R}
\right)
\left[
\frac{
\left( 
\hat{\rv}
\bcd
{\mathrm d}
\rv
\right)
\,
\hat{\rv}
\bmt
\hat{\xv}
}
{
r^2
\,
\left(
\hat{\rv}
\bcd
\hat{\xv}
+
1 
\right)
}
\right.
\nonumber
\\
&&
\left.
\left.
+
\left(
\frac
{1}
{r}
+
\frac
{1}
{|\xv|}
\right)
\,
\frac{ 
\left[
\left(
\hat{\rv} 
+
\hat{\xv}
\right)
\bmt
\left(
\hat{\rv}
\bmt
{\mathrm d}\rv
\right)
\right]
\bmt
\hat{\xv}
}
{
r
\,
\left(
\hat{\rv}
\bcd
\hat{\xv}
+
1 
\right)^2
}
\right]
\right\}
.
	\label{mmBSL0r}
\end{eqnarray}
The subphotospheric path $\Css$ here is the image of the coronal path $\Cc$ under the inversion transformation, which is defined point-wise by Equation (\ref{Rvs}).

One can see that there are two terms under the integral over the path $\Css$ in Equation (\ref{mmBSL0r}).
The first of them represents the classical Biot\textendash{}Savart field whose current, however, is ``modulated'' along the path by the inverse length of the radius vector $\Rv$.
Without such a modulation of the current along the path $\Css$,
the required vanishing of the total radial field at the boundary
cannot be reached, as previously demonstrated by Equation (\ref{RdBIKt}).

The second of the indicated terms describes the toroidal field generated by the line current flowing on the radial sides of magnetized infinitesimal triangles.
These triangles constitute the magnetic shell that is obtained by merging two original shells, which were set up on the ruled surfaces $\Sss$ and $\Ss$ with the same directrix $\Cs \equiv \Css$ (see Appendix \ref{A3} and Figure \ref{f:ScSs}).
The modulation factor
$
\left(
1
-
1
/
{\mathcal R}
\right)
$
here describes the fill and drain of the line current on the path $\Css$ by the currents flowing on the radial sides of the triangles (see Figure \ref{f:ScSs}(c)).
This factor vanishes at the foot points of $\Css$, where ${\mathcal R}=1$, which corresponds to the above-mentioned elimination of the toroidal field together with its source, i.e., the line current at the straight edges of $\Ss$.

Equation (\ref{Rvs}) allows one first to obtain the relationship between the line elements of the paths $\Css$ and $\Cc$ (see Equation (\ref{dRvs})) and then to reduce Equation (\ref{mmBSL0r}) to the following integral along the path $\Cc$:
\begin{eqnarray}
&&
\left[
\frac
{\mu I}
{4\pi R_{\sun}}
\right]
\qquad
\Bv_{I\ominus}
=
\int_{\Cc}
\left(
\frac
{
\hat{\rv}
}
{
r^2
}
-
\frac
{1}
{\mathcal R}
\frac
{
\hat{\rv}_{*}
}
{
\rs^2
}
\right)
\bmt
{\mathrm d}
\rv
+
\nonumber
\\
&&
\qquad
\left(
1
-
\frac
{1}
{\mathcal R}
\right)
\hat{\xv}
\bmt
\left[
\frac{
\hat{\rv}_{*}
\left( 
\hat{\rv}_{*}
\bcd
{\mathrm d}
\rv
\right)
}
{
\rs^2
\,
\left(
\hat{\rv}_{*}
\bcd
\hat{\xv}
+
1 
\right)
}
+
\left(
\frac
{1}
{\rs}
+
\frac
{1}
{|\xv|}
\right)
\,
\frac{ 
\left(
\hat{\rv}_{*} 
+
\hat{\xv}
\right)
\bmt
\left(
\hat{\rv}_{*}
\bmt
{\mathrm d}\rv
\right)
}
{
\rs
\,
\left(
\hat{\rv}_{*}
\bcd
\hat{\xv}
+
1 
\right)^2
}
\right]
,
	\label{mmBSL1r}
\end{eqnarray}
which is an alternative reduced form of the \MBSL{} formulated exclusively by means of the coronal-current path.

It is also useful to rewrite our \MBSL{} in terms of the vector potential.
We managed to calculate and transform it to the following compact expression:
\begin{eqnarray}
\left[
\frac
{\mu I}
{4\pi }
\right]
\qquad
\bm{A}_{I\ominus}
=
-
\int_{\Cc} 
\frac
{
{\mathrm d}
\rv
}
{
r
}
-
\int_{\Css}
\left[
\frac
{1}
{\mathcal R}
\frac
{
{\mathrm d}
\rv
}
{
r
}
+
\left(
1
-
\frac
{1}
{\mathcal R}
\right)
\hat{\xv}
\frac{
\left(
\hat{\rv}
+ 
\hat{\xv}
\right)
\bcd
{\mathrm d}
\rv
}
{
r
\left(
\hat{\rv}
\bcd
\hat{\xv}
+
1 
\right)
}
\right]
\!\!
.
	\label{AmmBSL0r}
\end{eqnarray}
Other than the ``modulated'' classical \BSL{} term, it has under the second integral a less-obvious term multiplied by
$
\left(
1
-
1
/
{\mathcal R}
\right)
$.
Its curl yields the term that enters into Equation (\ref{mmBSL0r}) with the same factor and, hence, describes the same magnetized infinitesimal triangles discussed above.
Therefore, it can be obtained by integrating the classical \BSL{} kernel $1/r$ of the vector potential over the radial sides of an infinitesimal triangle of the merged magnetic shell $\Sss$.
Alternatively, it can also be derived simply by setting
$
\Rv_{1}
=
\Rv
+
{\mathrm d}
\Rv
$
and
$
\Rv_{2}
=
\Rv
$
in Equation (\ref{AICs}) and then calculating the leading term of its Taylor expansion by
$
{\mathrm d}
\Rv
$.
Thus, the nontrivial term considered describes the toroidal part of the vector potential produced by each indicated triangle.

Using the same approach as in the derivation of Equation (\ref{mmBSL1r}), we obtain the alternative form for our vector potential formulated solely in terms of the coronal path $\Cc$:
\begin{eqnarray}
\left[
\frac
{\mu I}
{4\pi }
\right]
\qquad
\bm{A}_{I\ominus}
=
-
\int_{\Cc}
\left( 
\frac
{1}
{r}
-
\frac
{1}
{
{\mathcal R}
\rs
}
\right)
{\mathrm d}
\rv
+
\hat{\xv}
\int_{\Cc}
\left(
1
-
\frac
{1}
{\mathcal R}
\right)
\frac{
\left(
\hat{\rv}_{*}
+ 
\hat{\xv}
\right)
\bcd
{\mathrm d}
\rv
}
{
\rs
\left(
\hat{\rv}_{*}
\bcd
\hat{\xv}
+
1 
\right)
}
\,
.
	\label{AmmBSL1r}
\end{eqnarray}

Based on our previous studies (V. S. Titov et al. \citeyear{Titov2018}, \citeyear{Titov2021}), we now modify Equation (\ref{AmmBSL0r}) for a thin current channel with a circular cross section of radius $a$.
We replace $1/r$ in Equation (\ref{AmmBSL0r}) with $K_{I}(\tilde{r})/a$, where $K_{I}(\tilde{r})$ is a regularized \BSL{} kernel, a function of $\tilde{r}\equiv r/a$, which differs from zero for $\tilde{r}<1$ and smoothly transitions to $1/\tilde{r}$ for $\tilde{r}>1$.
Changing also
$
{\mathrm d}
\rv
$
for
$
-
\Rvp
{\mathrm d}
l
$,
we obtain the following generalized form of Equation (4) from \citet{Titov2021}:
\begin{eqnarray}
\left[
\frac
{\mu I}
{4\pi }
\right]
\quad
\bm{A}_{I}
=
\int_{\Cc} 
K_{I}
(\tilde{r})
\,
\Rvp
\frac
{
{\mathrm d}
l
}
{a}
+
\int_{\Css}
K_{I}
(\tilde{r})
\left[
\frac
{\Rvp}
{\mathcal R}
+
\left(
1
-
\frac
{1}
{\mathcal R}
\right)
\frac{
\left(
\hat{\rv}
+
\hat{\xv}
\right)
\bcd
\Rvp
}
{
\hat{\rv}
\bcd
\hat{\xv}
+
1 
}
\,
\hat{\xv}
\right]
\frac
{
{\mathrm d}
l
}
{a}
.
\nonumber
\\
	\label{AmmRBSL0}
\end{eqnarray}
Similarly to the above consideration of (\ref{AmmBSL0r}) and Equation (\ref{AmmBSL1r}), we can rewrite Equation (\ref{AmmRBSL0}) in terms of the line integral only along the coronal path $\Cc$:
\begin{eqnarray}
\left[
\frac
{\mu I}
{4\pi }
\right]
\quad
\bm{A}_{I}
=
\int_{\Cc}
\left\{
\left[
K_{I}
(\tilde{r})
-
\frac
{
K_{I}
(\tilde{r}_{*})
}
{
{\mathcal R}
}
\right]
\,
\Rvp
-
K_{I}
(\tilde{r}_{*})
\left(
1
-
\frac
{1}
{\mathcal R}
\right)
\frac{
\left(
\hat{\rv}_{*}
+ 
\hat{\xv}
\right)
\bcd
\Rvp
}
{
\hat{\rv}_{*}
\bcd
\hat{\xv}
+
1 
}
\,
\hat{\xv}
\right\}
\frac
{
{\mathrm d}
l
}
{a}
\,
.
\nonumber
\\
	\label{AmmRBSL}
\end{eqnarray}

A more detailed consideration and application of
this expression is beyond the scope of this paper and will be described in future work.
Here we just want to emphasize that, in our \RBSL{} method, it represents the axial vector potential of a thin current-unneutralized MFR.
By construction, the curl of $\bm{A}_{I}$ defines the azimuthal magnetic field of MFRs whose radial component vanishes at the photospheric boundary outside MFR footprints even if the length of a modeled MFR is comparable in value with the solar radius.
This new expression for $\bm{A}_{I}$ significantly extends the ability of the \RBSL{} method to model realistic MFR configurations.

\subsubsection{ Illustrating Example of the \MBSL{} Field
	\label{sss:mmBSLfex}}

\begin{figure}[ht!]
\centering
\resizebox{0.55\textwidth}{!}{
\includegraphics{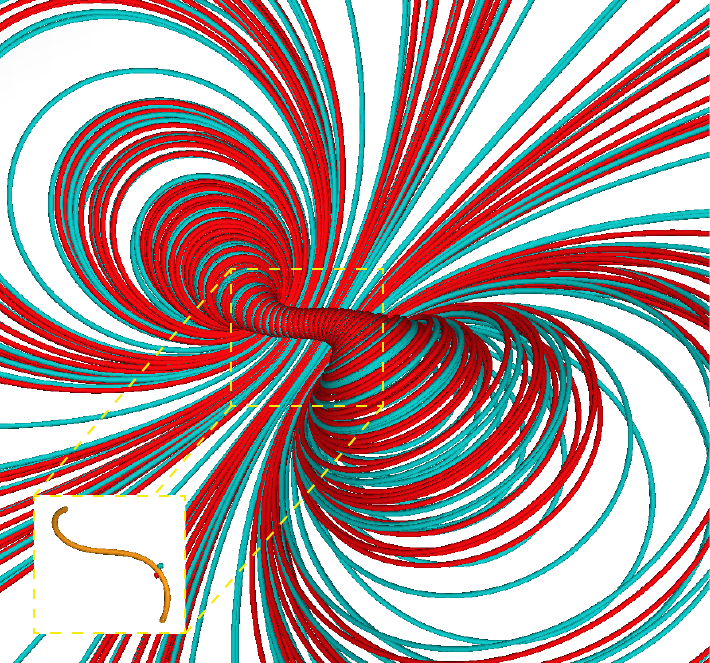}}
\caption{
Top view of magnetic field lines for the configuration defined by the magnetogram-matching Biot\textendash{}Savart law (Equation (\ref{mmBSL})); the inset shows the corresponding current path $\Cc$ (orange) and two arbitrary starting points (close to the path) for the plotted field lines.
	\label{f:mmBSL_FLs}
}
\end{figure}

Figure \ref{f:mmBSL_FLs} exemplifies $\bm{B}_{I\ominus}$ for a specific current path by showing two field lines whose lengths were restricted from above by $10^3 R_\sun$.
If plotted without any restriction in length, these field lines would indefinitely fill a part of the coronal volume, since they are disconnected from the boundary.
However, the superposition of $\bm{B}_{I\ominus}$ on, say, a bipolar ambient potential magnetic field would ``short circuit'' the previously disconnected field lines to the photosphere.

Despite the abstract meaning of $\bm{B}_{I\ominus}$, its structure helps us to understand the origin of some observed morphological features, such as hook-like loops in sigmoidal MFR configurations.
Indeed, the modeling of such configurations suggests that the MFR footprints reside at the periphery of magnetic flux spots (\citeauthor{Titov2018} \citeyear{Titov2018}, \citeyear{Titov2021}).
This means that, in our composite configurations,
$\bm{B}_{I\ominus}$ would likely prevail over the ambient field near foot points of the current path,
so large curls of the field lines at these places (see Figure \ref{f:mmBSL_FLs}) should approximately track the corresponding field lines in the sigmoidal configurations.
Being ``short-circuited'' by a weak but nonvanishing ambient field, these curls acquire the shape of hooks that embrace the current path near its foot points.
In other words, these features are formed due to a nonzero current flowing in a narrow channel whose footprints are located in weak-field regions.
Thus, they can be considered as indirect evidence that the current in the MFRs of sigmoidal configurations is unneutralized.

\subsection{Magnetic Configurations with Distributed Currents
	\label{ss2:mmBSl}
}

The results described in Section \ref{ss1:mmBSl} for a closed-line current can
straightforwardly be extended to realistic configurations, where the current is distributed in the coronal volume.
To this end, we approximate the distributed current as running through a continuum of wires with infinitesimal cross sections.
The individual contribution of the wire to the total field,
the photospheric $B_r$ of which vanishes, can be obtained by modifying Equation (\ref{mmBSL}) as shown below.

We find it convenient for applications to derive the \MBSL{} for distributed currents in a dimensional form, where the current density $\Jv$ is measured in $A/m^2$, while all space vectors and lengths are still normalized to $R_{\sun}$ as in our consideration above. Having this in mind, we should change $\mathrm{d}\rv$ in Equation (\ref{mmBSL}) for $-\Jv\, \mathrm{d}V$,
where the volume element 
$
\mathrm{d}
V
\equiv
\mathrm{d}^3
\bm{R}
$
(normalized to $R_\sun^3$) refers to a point $\Rv\equiv\bm{R}$.
By doing this substitution, we just follow a similar generalization of the classical \BSL{} for a wire to the system with a volumetrically distributed current.

The contribution of the subphotospheric closure current described by $\bm{\Theta}(\xv,\rv)$ must be weighted with the corresponding total current
{
$
(
\Jv
\bcd
\Rvu
)
\,
\mathrm{d}
\Omega
$}
in the wire, where
{
$
\mathrm{d}
\Omega
\equiv
\mathrm{d}^2
\Rvu
$}
is the boundary-surface element or the increment of the solid angle $\Omega$ and $\Rvu$ is the normal to the boundary at the first foot point, $\Rv_1=\bm{R}=\Rvu$, of the wire.
The second of the two conjugate foot points should be left
out of this expression, as we are going to sum up their contributions point by point rather than by pairs.

By superimposing the contributions of all of these wires to the total field with the help of Equations (\ref{dBI}), (\ref{dBC}), and (\ref{Thev}), we arrive at the following dimensional form of the \MBSL{} 
for a distributed current in the coronal volume
$V$
$
\left(
\bm{R}
\in
V
:
|\bm{R}|
\ge
1
\right)
$
with the photospheric boundary
$\partial V$
$
\left(
\bm{R}
\in
\partial V
:
|\bm{R}|
=
1
\right)
$
:
\begin{eqnarray}
&&
\Btld
=
\frac
{\mu R_{\sun}}
{4\pi}
\int_{V}
{\mathrm d}
V
\left\{
\Jv
\bmt
\left(
\frac
{
\hat{\rv}
\,
}
{
r^2
}
-
\frac
{
\hat{\rv}_{*}
}
{
R
r_*^2
}
\right)
\right.
\nonumber
\\
&&
\qquad
-
\frac
{1}
{R}
\left[
\frac{
\left( 
\hat{\rv}_{*}
\bcd
\Jv
\right)
\hat{\rv}_{*}
}
{
\rs^2
\,
\left(
\hat{\rv}_{*}
\bcd
\hat{\xv}
+
1 
\right)
}
+
\left.
\left(
\frac
{1}
{\rs}
+
\frac
{1}
{|\xv|}
\right)
\,
\frac{
\left(
\hat{\rv}_{*} 
+
\hat{\xv}
\right)
\bmt
\left(
\hat{\rv}_{*}
\bmt
\Jv
\right)
}
{
\rs
\,
\left(
\hat{\rv}_{*}
\bcd
\hat{\xv}
+
1 
\right)^2
}
\right]
\bmt
\hat{\xv}
\right\}
\nonumber
\\
&&
\qquad
+
\,
\hat{\xv}
\bmt
\frac
{\mu R_{\sun}}
{4\pi}
\int_{\partial V}
{\mathrm d}
\Omega
\left(
\frac
{1}
{r}
+
\frac
{1}
{|\xv|}
\right)
\frac{
(
\Jv
\bcd
\Rvu
)
\,
\hat{\rv}
}
{
\hat{\rv}
\bcd
\hat{\xv}
+
1
}
\,
,
	\label{mmBSL2}
\end{eqnarray}
where
$
\rv
=
\xv
-
\bm{R}
$
is the vector from a volume or surface element located at the point $\bm{R}$ to the observation point $\xv$, $\Jv$ is the dimensional density of the current at the point $\bm{R}$, and $\hat{\rv} \equiv \rv/r$.
Similarly,
$
\rv_*
=
\xv
-
\bm{R}_*
$
is the vector from the inversion image point $\bm{R}_*=\bm{R}/R^2$ to the observation point $\xv$, and $\hat{\rv}_{*} \equiv \rv_{*}/r_{*}$.

{
Let us express the surface term in Equation (\ref{mmBSL2}) by means of} the vector potential $\bm{A}_{f}$ to make it obvious that this term describes a toroidal magnetic field.
For this purpose, the part of Equation (\ref{AICs}) that refers to the first foot point, $\Rv_1=\bm{R}$, of the wire should be weighted with the total current
$
(
\Jv
\bcd
\Rvu
)
\,
\mathrm{d}
\Omega
$
in the wire and then integrated over the boundary surface.
Using explicitly Equation (\ref{Txr}) in this integral, one obtains the following result:
\begin{eqnarray}
\bm{A}_{f}
&=&
f(\xv)
\,
\hat{\xv}
\,
,
	\label{ATv}
\\
f(\xv)
&=&
\mu
R_{\sun}^2
\int_{\partial V}
\mathrm{d}\Omega
\,
(
\Jv
\bcd
\Rvu
)
\,
G_{f}
\,
,
	\label{TFFF}
\\
G_{f}
&=&
-
\frac
{1}
{4\pi}
\ln
\left(
\xv
\bcd
\rv
+
|\xv|
r
\right)
\,
,
	\label{GT1}
\end{eqnarray}
which shows that the TFFF, $f(\xv)$, is determined via the convolution of the photospheric radial component of the current density
$
(
\Jv
\bcd
\Rvu
)
$
with the function $G_f$ given by Equation (\ref{GT1}).
In this respect, $G_f$ is similar to the Green function.
However, the calculation of the radial current density corresponding to $G_f$ shows that its spike is not represented by the Dirac delta function, as it would be for the true Green function.
Therefore, we will call $G_f$ the {\it source function}.

By construction, the surface term in Equation (\ref{mmBSL2}) describes the toroidal field related only to subphotospheric closure currents.
The other part of this field, both in the volume and at the surface, is produced by coronal currents.
As will be explained in Section \ref{s:dvm}, the total photospheric toroidal field is generated by all elements of the current tubes that penetrate the surface.

Generalizing Equation (\ref{mmBSL}) to Equation (\ref{mmBSL2}), 
we used as an idealized prototype the configuration with a line current rooted at the surface.
This seems, at first, to imply that our generalized configuration with a distributed current should consist of only elementary current tubes that are also rooted at the surface.
However, this implication is not correct, and Equation (\ref{mmBSL2}) actually describes the configurations with a more general current connectivity.
Indeed, the elementary \MBSL{} fields are integrated in Equation (\ref{mmBSL2}) elements-wise throughout the exterior volume, the current elements of which contribute to the integral identically, regardless of whether or not they provide coronal closure to the interior currents piercing the surface.
In particular, a set of these elements can form elementary current tubes that are fully disconnected from the surface, as occurs, for example, in the heliospheric current layer.
The presence of the disconnected current tubes does not affect, at the surface, the radial components of both magnetic and current-density fields, which is sufficient then for Equation (\ref{mmBSL2}) to be applicable to such configurations.

However, the connectivity of the current elements does matter as to whether or not the resulting field at the surface is potential.
As follows from Section (\ref{sss:eBSLnonpot}), only the elements that form the current tubes that are closed and fully detached from the surface generate a potential field in it.
In contrast, current tubes that are rooted, with at least one of their ends at the surface, make the photospheric field nonpotential.
The latter is also true, of course, for the coronal-current tubes touching the surface, even if these tubes are partly detached from it and form closed circuits. 
This would imply the presence of a nonvanishing tangential current density on the upper side of this surface, i.e., at $|\xv| = 1_{+0}$, which means the nonpotentiality of the magnetic field there.

In this respect, it looks, at first sight, inconsistent that \citet{Schuck2022} postulated that the two poloidal fields in their decomposition
must be purely potential beyond the surface, one above and the other below it.
This postulate is substantiated by invoking Gauss's insight on
terrestrial toroidal currents, namely that similar currents flowing separately in the solar interior and exterior can generate only a potential poloidal field at the upper and lower sides of the boundary surface, at $|\xv| = 1_{+0}$ and $|\xv| = 1_{-0}$, respectively.
As already mentioned in Section \ref{s:intro}, for the photospheric
conditions, however, 
it is better to have all decomposed parts determined at the same surface side, preferably at $|\xv| = 1_{+0}$.
The latter can be achieved by simply reassigning $\BPgt$ to this level
if the current density associated with $\BPgt$ is {\it continuous above the surface}.
This is because a cross-surface jump of this current density
can cause only a jump in the radial derivatives of $\BPgt$, but not in its values.
Therefore, the above-mentioned inconsistency of the Gaussian decomposition is only apparent.
In Appendix \ref{s:relation} we consider an example of a simple magnetic configuration that illustrates this point.

Following a similar approach to the one we used to derive Equation (\ref{mmBSL1r}) for the line current, we obtain an alternative form of \MBSL{} for distributed currents, which is expressed solely in terms of the integral over the coronal volume:
\begin{eqnarray}
&&
\Btld
=
\frac
{\mu R_{\sun}}
{4\pi}
\int_{V}
{\mathrm d}
V
\left\{
\Jv
\bmt
\left(
\frac
{
\hat{\rv}
\,
}
{
r^2
}
-
\frac
{
\hat{\rv}_{*}
}
{
R
r_*^2
}
\right)
\right.
\nonumber
\\
&&
\qquad
+
\left(
1
-
\frac
{1}
{R}
\right)
\left[
\frac{
\left( 
\hat{\rv}_{*}
\bcd
\Jv
\right)
\hat{\rv}_{*}
}
{
\rs^2
\,
\left(
\hat{\rv}_{*}
\bcd
\hat{\xv}
+
1 
\right)
}
\left.
+
\left(
\frac
{1}
{\rs}
+
\frac
{1}
{|\xv|}
\right)
\,
\frac{
\left(
\hat{\rv}_{*} 
+
\hat{\xv}
\right)
\bmt
\left(
\hat{\rv}_{*}
\bmt
\Jv
\right)
}
{
\rs
\,
\left(
\hat{\rv}_{*}
\bcd
\hat{\xv}
+
1 
\right)^2
}
\right]
\bmt
\hat{\xv}
\right\}
.
\nonumber
\\
	\label{mmBSLc}
\end{eqnarray}

The magnetic field of coronal currents would look exactly like $\Btld$ if the solar globe were an ideal rigid conductor with the surface polarization currents that completely shield the globe interior from penetration of the coronal magnetic flux.
In our \MBSL{} representation, the role of such polarization currents is played by subphotospheric closure currents and elementary fictitious magnetic shells described in Section \ref{sss:dBCdBCs} and Appendix \ref{s:dBC}.

We expect that the form of \MBSL{} given by Equation (\ref{mmBSLc}) can be useful for
producing NLFFF extrapolations in spherical geometry, as it directly determines $\Btld$ from a given distribution of $\Jv$ in real space.
To find a similar relationship between $\Btld$ and $\Jv$, \citet{Gilchrist2014} used a global representation of the magnetic field with vector spherical harmonics.
The calculation of this relationship in real space can
provide more clarity, which we used above
when generalizing our \RBSL{} for modeling elongated MFRs (see Equation (\ref{AmmRBSL})).
It is not difficult to verify that in the limit of vanishing curvature of the solar surface, only the first term in the integrand of Equation (\ref{mmBSLc}) survives.
This term represents coronal-current elements and their images that are mirrored about the surface, exactly as required for keeping $\tilde{B}_r=0$ in planar geometry.

\section{Decomposition of Vector Magnetograms
	\label{s:dvm}}

Section \ref{s:mmBSl} demonstrates that the subphotospheric currents manifest themselves in the corona exclusively through the photospheric distributions of the radial magnetic field $B_r$ and current density $J_r$---no other parameters related to the interior currents affect the exterior magnetic fields.
According to our \MBSL{} approach, the photospheric magnetic field, as defined at $|\xv|=1_{+0}$, can be decomposed into the following three parts:
\begin{enumerate}

\item 
The potential magnetic field $\Bv_{\mathrm{pot}}$ whose $B_r$ is generated by subphotospheric currents that do not flow beyond the surface $|\xv|=1$.

\item
The toroidal magnetic field, which is superposed of the \BSL{} fields produced by subphotospheric closure currents and fictitious magnetic shells compensating $B_r$ of those currents;
the resulting field is determined solely in terms of the photospheric $J_r$ distribution and does not depend on the paths of the closure currents.

\item
The magnetic field generated by coronal currents under the condition that their photospheric $B_r$ vanishes; this condition is sustained by additional fictitious subphotospheric magnetic shells.

\end{enumerate}
In our previous preliminary study \citep{Titov2024X} we demonstrated that this decomposition allows one to identify the location of MFRs in projection to the photospheric surface.
It is particularly important that such a localization of MFRs can be done in advance of modeling PECs by using only magnetic data.

However, we have recently realized that there is one aspect of this decomposition that is not fully satisfactory, namely, that its third part, associated with the coronal currents, includes both poloidal and toroidal components.
In a more consistent decomposition, the
toroidal and poloidal fields should be separated from each other.
Fortunately, the corresponding redistribution of these fields within our decomposition is not difficult to perform.

Indeed, the total coronal magnetic field is
\begin{eqnarray}
\Bv
=
\Btld
+
\Bv_{\mathrm{pot}}
\,
,
	\label{Bv}
\end{eqnarray}
where $\Btld$ is our \MBSL{} field, which is equivalently described by either Equation (\ref{mmBSL2}) or Equation (\ref{mmBSLc}).
Using Equations (1b) and (6c) from \citet[][]{Schuck2022} and our Equation (\ref{Bv}), we see that the total toroidal field $\Bv_{T}$ on the surface in our approach is described as follows:
\begin{eqnarray}
&&
\underline{
\Bv_{T}
=
\nt T
\bmt
\hat{\xv}
}
\,
,
	\label{BTdef}
\\
&&
\hat{\xv}
\bcd
\bm{\nabla}
\bmt
\Bv
=
\hat{\xv}
\bcd
\nt
\bmt
\Btld
=
\hat{\xv}
\bcd
\nt
\bmt
\Bv_{T}
\nonumber
\\
&&
=
\underline{
-
\nt^2
T
=
\mu
R_{\sun}
(
\hat{\xv}
\bcd
\Jv
)
}
\,
.
	\label{Teqs}
\end{eqnarray}
where $T$ is the toroidal scalar field or the TFFF of the total toroidal field, and
\begin{eqnarray*}
\nt = \bm{\nabla}
-
\hat{\xv}
\,
(
\hat{\xv}
\bcd
\bm{\nabla}
)
\end{eqnarray*}
is the dimensionless operator $\bm{\nabla}$ that acts tangentially to our boundary surface $|\xv|=1$.

Although, by construction, $\Btld$ is strictly tangential to the boundary, its surface divergence can generally differ from zero.
This is because vanishing of its radial component,
$
(
\hat{\xv}
\bcd
\Btld
)
\bigr|
_{|\xv|=1_{+0}}
$,
does not imply that the radial derivative of this component,
$
(
\hat{\xv}
\bcd
\bm{\nabla}
)
(
\hat{\xv}
\bcd
\Btld
)
\bigr|
_{|\xv|=1_{+0}}
$,
also vanishes.
Using this fact and Equations (1a) and (6b) from \citep[][]{Schuck2022}, we obtain the remaining photospheric poloidal part of
$
\Btld
$, that is,
\begin{eqnarray}
\Btld
-
\Bv_{T}
\equiv
\underline{
\Bv_{\St}
=
\nt
\St
}
\,
,
	\label{BSdef}
\end{eqnarray}
the following relationships:
\begin{eqnarray}
&&
\nt
\bcd
\Bv_{\tilde{S}}
=
-
(
\hat{\xv}
\bcd
\bm{\nabla}
)
(
\hat{\xv}
\bcd
\Bv_{\tilde{S}}
)
=
\underline{
\nt^2
\tilde{S}
=
\nt
\bcd
\Btld
}
\,
,
	\label{Seqs}
\end{eqnarray}
where
$
\tilde{S}
$
is the spheroidal scalar field for
$
\Btld
\bigr|
_{|\xv|=1_{+0}}
$
or simply the surface potential for the tangential poloidal field
$
\Bv_{\tilde{S}}
\bigr|
_{|\xv|=1_{+0}}
$.
The tilde symbol is used in $\St$ to emphasize that this potential does not describe the full poloidal field but only a part of it, the one that is associated only with currents flowing in the corona,
regardless of their connectivity to the boundary.

Thus, by combining Equations (\ref{Bv}) and (\ref{BSdef}),  we arrive at the desired decomposition of a given photospheric magnetic field $\Bv$ at $|\xv|=1_{+0}$:
\begin{eqnarray}
\boxed{
\Bv
=
\Bv_{\mathrm{pot}}
+
\Bv_{T}
+
\Bv_{\tilde{S}}
}
\,
,
	\label{decomp2}
\end{eqnarray}
in which the potential magnetic field
$
\Bpot
$
can be calculated in a common way by using $B_{r}$ of the measured magnetograms, so that $B_r$ is identical to the radial component of $
\Bpot
$.
The toroidal and poloidal fields, $\Bv_{T}$ and $\Bv_{\tilde{S}}$, both purely tangential to the surface, are defined in terms of their scalar fields, $T$ and $\tilde{S}$, by Equations (\ref{BTdef}) and (\ref{BSdef}), respectively. 
These scalar fields, in turn, are solutions of the corresponding Poisson's equations on the sphere $|\xv|=1$, which are Equations (\ref{Teqs}) and (\ref{Seqs}), the right-hand sides of which are defined by the surface curl and divergence, respectively, of the field
$
\Btld
=
\Bv
-
\Bv_{\mathrm{pot}}
$,
which itself is derived from vector magnetograms.

In principle, one can determine
$
\Bv_{\tilde{S}}
$
without first calculating the potential $\tilde{S}$ and then its surface gradient: it can be done simply by combining Equations (\ref{Bv}) and (\ref{BSdef}) as follows:
\begin{eqnarray}
\Bv_{\tilde{S}}
=
\Bv
-
\Bv_{\mathrm{pot}}
-
\Bv_{T}
\,
.
	\label{BSdef2}
\end{eqnarray}
However, as will be clarified later, it is still worth calculating the potential $\tilde{S}$, because the representation of
$\BvSt$
in terms of $\tilde{S}$ has its own merits.
Equation (\ref{BSdef2}) then can be used to validate that it yields the same result as Equation (\ref{BSdef}).

Our poloidal field $\BvSt$ is defined at the same level, $|\xv|=1_{+0}$, as our other decomposed parts of the field. 
It has no radial component and is potential as a surface vector field,
but generally {\it nonpotential} as a 3D vector field.
This is because, in general, there is a nonvanishing radial gradient of $\BvSt$
at $|\xv|=1_{+0}$, which sustains the corresponding toroidal current density there.

In this respect, it seems paradoxical that, similarly to the toroidal field $\BvT$, the poloidal field $\BvSt$ is a purely tangential field.
However, this apparent paradox is resolved if we employ our \MBSL{} representation of a current-carrying 3D field.
In accordance with this representation,
the field $\BvT + \BvSt$ is produced by real (coronal) and fictitious (interior) currents, which sustain $B_r=0$ on the surface.
In particular, the field $\BvSt$ is generated by coronal toroidal currents and their ``mirror'' counterparts below the surface, so that the resulting $\BvSt$ has no radial component.
We place the term mirror in quotes to indicate that this term actually stands for
our generalization of this concept to spherical geometry, which is described in Section \ref{s:mmBSl}.

Compared to the decomposition by \citet{Schuck2022}, our decomposition is defined by Equation (\ref{decomp2}) and also has three parts, one of which, $\Bv_{T}$, is identical in both decompositions, while the other two parts are quite different.

Our potential magnetic field $\Bv_{\mathrm{pot}}$ refers to the level $|\xv|=1_{+0}$, i.e., to the upper side of the surface, and is uniquely determined by the observed photospheric $B_{r}$ distribution.
The analogous potential poloidal field $\BPlt$ in \citep{Schuck2022} refers to the same level, but corresponds only to a part of this distribution.
Their other potential poloidal field $\BPgt$ corresponds to the remaining part of the observed $B_{r}$ distribution and, by construction, refers to the lower side of the surface, that is, strictly speaking, to the different level $|\xv|=1_{-0}$.
However, as mentioned in Sections 1 and 2 and illustrated in Appendix \ref{s:relation}, for a continuous distribution of the current density above the surface, the poloidal field $\BPgt$ must
have the same values on both sides of the surface.
Therefore, we are allowed to simply raise the original values of $\BPgt$ to the level $|\xv|=1_{+0}$ while keeping $\BPlt$ unchanged.
The same infinitesimal lift is obviously valid for the toroidal field $\BvT$, which was originally defined at $|\xv|=1$.
Thus, all three parts of the Gaussian method
can be defined at the same level as in our method, which means that both decompositions are complete under the continuity condition of the current density.
Mathematically, the difference between them is only in how the whole poloidal part of the photospheric field is partitioned.
It has yet to be seen whether this difference and the possible complementarity of the two methods are
important in practice.
Regardless of this question, our method shows great potential for extending the analysis of vector magnetic data, as illustrated in Section \ref{examples}.

\subsection{Potential Magnetic Field at the Boundary}

Let longitude $\phi$ and colatitude $\theta$ represent an observation point $\hat{\xv}$ at the photospheric boundary.
In the global Cartesian system of coordinates with the origin at the center of the Sun, we have
\begin{eqnarray}
\hat{\xv}
=
\left(
\sin\theta
\,
\cos\phi
,
\sin\theta
\,
\sin\phi
,
\cos\theta
\right)
\,
.
	\label{xhv}
\end{eqnarray}
Similarly, the unit vector
\begin{eqnarray}
\Rvu
=
\left(
\sin\theta^{\prime}
\cos\phi^{\prime}
,
\sin\theta^{\prime}
\sin\phi^{\prime}
,
\cos\theta^{\prime}
\right)
	\label{Rhv}
\end{eqnarray}
then represents a source point with longitude $\phi^{\prime}$ and colatitude $\theta^{\prime}$, where we assume the radial magnetic component,
$
B_{r}
\equiv
(
\Bv
\bcd
\Rvu
)
$,
to be known.
To derive the expression for the photospheric tangential component of the potential magnetic field, we will use the Green function for the external Neumann problem of the Laplace equation in spherical geometry.
\citeauthor{Nemenman1999} (\citeyear{Nemenman1999}; see their Equation (8)) provided an explicit formula for this Green function.
For our length normalization and chosen notation of variables, the latter can be written as follows: 
\begin{eqnarray}
G_\mathrm{pot}
=
\frac
{1}
{4\pi}
\left[
\frac
{2}
{
|
\xv
-
\Rvu
|
}
-
\ln
\left(
\frac
{
1
-
\xv
\bcd
\Rvu
+
|
\xv
-
\Rvu
|
}
{
|\xv|
-
\xv
\bcd
\Rvu
}
\right)
\right]
\,
.
\nonumber
\\
	\label{GP}
\end{eqnarray}

Having been multiplied by magnetic flux
$
(
\Bv
\bcd
\Rvu
)
R_{\sun}^2
\,
\mathrm{d}
\Omega^{\prime}
$
at a source point $\Rvu$,
where
$
\mathrm{d}
\Omega^{\prime}
=
\sin\theta^{\prime}
\,
\mathrm{d}
\theta^{\prime}
\,
\mathrm{d}
\phi^{\prime}
$
is an increment of the solid angle at this point, this function defines the corresponding contribution of the source to the scalar magnetic potential at a given observation point $\xv$.
Therefore, the convolution over the unit sphere,
\begin{eqnarray}
\Bv_\mathrm{pot}
=
-
\int
\mathrm{d}\Omega^{\prime}
\,
(
\Bv
\bcd
\Rvu
)
\bm{\nabla}_{\xv}
G_\mathrm{pot}
\,
,
	\label{BP}
\end{eqnarray}
defines the total potential magnetic field $\Bv_\mathrm{pot}(\xv)$ at $|\xv|>1$  produced by all photospheric sources.

Eliminating now the radial component from this expression, we obtain at $\xv \rightarrow \hat{\xv}$ the following formula for the tangential potential field on the surface:
\begin{eqnarray}
&&
\hat{\xv}
\bmt
(
\Bv_\mathrm{pot}
\bmt
\hat{\xv}
)
\est{|\xv|=1_{+0}}
=
\int
\mathrm{d}\Omega^{\prime}
\,
(
\Bv
\bcd
\Rvu
)
\,
\bm{G}_{B_{\mathrm{pot}}}
\,
,
\nonumber
\\
	\label{BPt}
\\
&&
\bm{G}_{B_{\mathrm{pot}}}
\equiv
-
\nt
G_\mathrm{pot}
\est{|\xv|=1_{+0}}
=
-
\frac
{
(
\et
\bcd
\Rvu
)
\,
\et
+
(
\ep
\bcd
\Rvu
)
\,
\ep
}
{
\pi
\left(
2
+
\Delta
\right)
\Delta^{3}
}
,
	\label{GPt}
\end{eqnarray}
where $\et$ and $\ep$ are the corresponding unit vectors of our spherical coordinate system at the observation point $\hat{\xv}$ and
\begin{eqnarray}
\Delta
&=&
\sqrt{2}
\,
(
1
-
\hat{\xv}
\bcd
\Rvu
)
^{1/2}
\,
	\label{Delta}
\end{eqnarray}
is the length of the chord that connects the source and observation points.
Trigonometric calculations yield
\begin{eqnarray}
\hat{\xv}
\bcd
\Rvu
&=&
\frac{1}{2}
\left\{
\left[
1
+
\cos
(
\phi
-
\phi^{\prime}
)
\right]
\cos
(
\theta
-
\theta^{\prime}
)
\right.
\nonumber
\\
&+&
\left.
\left[
1
-
\cos
(
\phi
-
\phi^{\prime}
)
\right]
\,
\cos
(
\theta
+
\theta^{\prime}
)
\right\}
\,
,
	\label{xR}   
\\
\ep
\bcd
\Rvu
&=&
-
\sin
\theta^{\prime}
\,
\sin
(
\phi
-
{\phi}
)
\, ,
	\label{phR}
\\
\et
\bcd
\Rvu
&=&
\frac
{1}
{2}
\left[
\sin
(
\theta
+
\theta^{\prime}
)
-
\sin
(
\theta
-
\theta^{\prime}
)
\right]
\cos
(
\phi
-
\phi^{\prime}
)
\nonumber
\\
&&
-
\frac
{1}
{2}
\left[
\sin
(
\theta
+
\theta^{\prime}
)
+
\sin
(
\theta
-
\theta^{\prime}
)
\right]
\, .
	\label{thR}
\end{eqnarray}

\subsection{Toroidal Magnetic Field at the Boundary}

Assume that the surface distribution of the radial component of the current density,
$
J_{r}
\equiv
(
\Jv
\bcd
\Rvu
)
$,
is known at each point $\Rvu$ of the surface.
Then the toroidal scalar field $T$, or the full TFFF, is a solution of Poisson's equation on the unit sphere, which is defined by the underlined part of Equation (\ref{Teqs}).
This solution can be represented by the convolution of this distribution with the corresponding Green function \citep[see, e.g.,][]{Beltran2019}
\begin{eqnarray}
G_{\mathrm{SP}}
&=&
-
\frac
{1}
{2\pi}
\ln
\Delta
\,
,
	\label{GSP}
\end{eqnarray}
as follows:
\begin{eqnarray}
T
=
\mu
R_{\sun}
\int
\mathrm{d}\Omega^{\prime}
\,
(
\Jv
\bcd
\Rvu
)
\,
G_{\mathrm{SP}}
\,
.
	\label{Tcon}
\end{eqnarray}

Application of this formula to particular $J_r$ distributions shows that the resulting $T$ is approximately twice as large as the incomplete TFFF, 
$
f
\bigr|_{|\xv|=1_{+0}}
$,
 defined by Equation (\ref{TFFF}) and produced only by subphotospheric closure currents.
This result is expected, as the source function, given by Equation (\ref{GT1}), on the surface transforms to
\begin{eqnarray}
G_{f}
\est{|\xv|=1_{+0}}
=
-
\frac
{1}
{4\pi}
\ln
\left[
\Delta
\left(
1
+
\Delta
/
2
\right)
\right]
\,
,
	\label{Gfnorm}
\end{eqnarray}
which approximately equals $G_{\mathrm{SP}}/2$  at small $\Delta$, where the main contributions to the convolution come from.

Using now Equations (\ref{BTdef}) and (\ref{Tcon}), we obtain the following expression for the full toroidal vector field itself on the surface:
\begin{eqnarray}
\Bv_{T}
&=&
\mu
R_{\sun}
\int
\mathrm{d}\Omega^{\prime}
\,
(
\Jv
\bcd
\Rvu
)
\,
\bm{G}_{B_{T}}
\,
,
	\label{BTcon}
\\
\bm{G}_{B_{T}}
&\equiv&
\nt
G_{\mathrm{SP}}
\bmt
\hat{\xv}
=
\frac
{
1
}
{
2
\pi
\Delta^2
}
\left[
(
\ep
\bcd
\Rvu
)
\,
\et
-
(
\et
\bcd
\Rvu
)
\,
\ep
\right]
\,
,
	\label{GBT}
\end{eqnarray}
where Equations (\ref{Delta})\textendash(\ref{thR}) should be applied to completely specify
$
\bm{G}_{B_{T}}
$.

Thus, both the TFFF and the toroidal magnetic field can be determined at the boundary through the convolutions of the surface radial current density,
$
(
\Jv
\bcd
\Rvu
)
$,
and the corresponding Green functions.

\subsection{Current-Carrying Poloidal Field at the Boundary
	\label{ss:BS}}

The spheroidal scalar field $\tilde{S}$ is also a solution of Poisson's equation on the sphere $|\xv|=1$, which is defined by the underlined part of Equation (\ref{Seqs}).
The right-hand side of this equation is the surface divergence of the tangential field
$
\Btld
=
\Bv
-
\Bv_{\mathrm{pot}}
$,
which is derived from Equations (\ref{BPt}) and (\ref{GPt}), and the corresponding vector magnetic data.
Thus, for a given divergence of $\Btld$, we obtain
\begin{eqnarray}
\St
=
-
\int
\mathrm{d}\Omega^{\prime}
\,
(
\nt
\bcd
\Btld
)
\,
G_{\mathrm{SP}}
\,
,
	\label{Scon}
\end{eqnarray}
where $G_{\mathrm{SP}}$ is defined by Equation (\ref{GSP}).

Following the definition of the poloidal field, which is provided by the underlined part of Equations (\ref{BSdef}), we now take the surface gradient of Equation (\ref{Scon}) to obtain
\begin{eqnarray}
\Bv_{\St}
&=&
\int
\mathrm{d}\Omega^{\prime}
\,
(
\nt
\bcd
\Btld
)
\,
\bm{G}_{B_{\St}}
\,
,
	\label{BScon}
\\
\bm{G}_{B_{\St}}
&\equiv&
\nt
G_{\mathrm{SP}}
=
\frac
{
1
}
{
2
\pi
\Delta^2
}
\left[
(
\et
\bcd
\Rvu
)
\,
\et
+
(
\ep
\bcd
\Rvu
)
\,
\ep
\right]
,
	\label{GBS}
\end{eqnarray}
where Equations (\ref{Delta})\textendash(\ref{thR}) should be used again to fully specify
$
\bm{G}_{B_{\St}}
$.

Thus, both the spheroidal potential $\St$ and the corresponding tangential poloidal field
$
\Bv_{\St}
$ are determined at the boundary via the convolutions of the surface divergence
$
\nt
\bcd
\Btld
$
and corresponding Green functions.

\subsection{Examples of Vector Magnetogram Decomposition
	\label{examples}}

\begin{figure*}[ht!]
\centering
\resizebox{0.99\textwidth}{!}{
\includegraphics{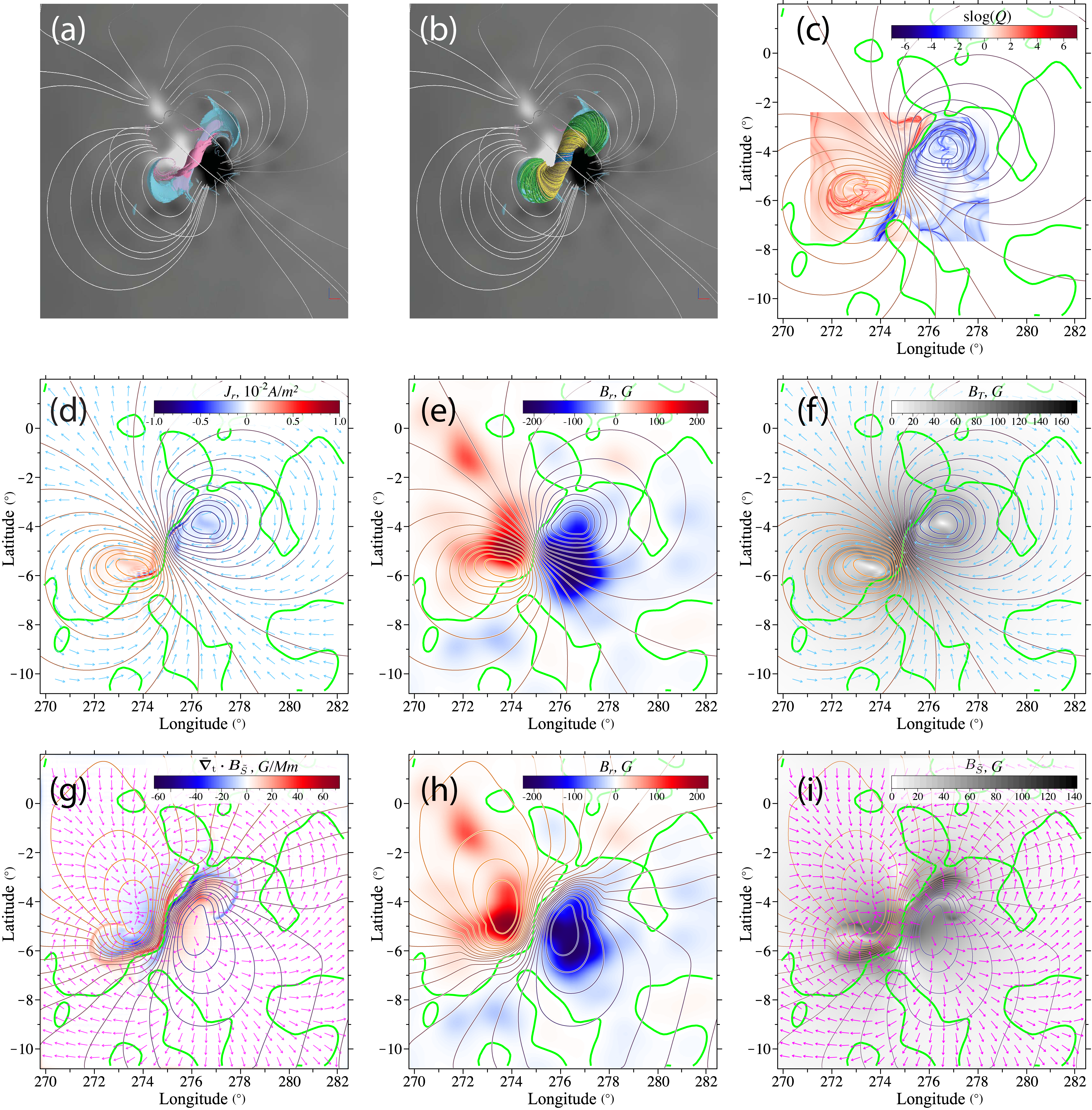}}
\caption{ The decomposition of the photospheric magnetic field for the PEC model of the 2009 February 13 CME:
{
 top view on the current (a) and magnetic field (b) structures \citep[see details in][]{Titov2021}; iso-contours of the toroidal scalar field $T$ are superimposed on the photospheric distributions of $\slog Q$ (c), $J_r$ (d), $B_r$ (e),  toroidal field $B_T$ (f);
iso-contours of the spheroidal scalar field $\tilde{S}$ are superimposed on the photospheric distributions of the surface divergence $\bar{\bm{\nabla}}_{\mathrm{t}}\bcd \BvSt \equiv \nt \bcd \BvSt / R_{\sun}$ (g), $B_r$ (h), and poloidal field $B_{\tilde{S}}$ (i);
both types of iso-contours are equally spaced, and their colors stepwise change from dark blue to light brown with growing the corresponding values of $T$ and $\tilde{S}$; cyan and magenta arrows, which are set on a grid of equidistant points, depict the directional fields of $\Bv_T$ (c) and $\Bv_{\tilde{S}}$ (f), respectively;
the thick green line represents the PIL.
}
\label{Feb13_Bph_dec}}
\end{figure*}

To see how our decomposition method can help in analyzing observed vector magnetograms, note first that iso-contours of the toroidal scalar field $T$, defined by Equation (\ref{Tcon}), represent the field lines of
$
\Bv_T
$,
defined by Equations (\ref{BTcon}) and (\ref{GBT}).
Therefore, plotting equally spaced iso-contours of $T$ and superimposing them on the corresponding distribution of
$
B_{T}
\equiv
|
\Bv_{T}
|
$
is a natural way to visualize the toroidal field on the surface.

Similarly, iso-contours of the spheroidal potential $\St$, defined by Equation (\ref{Scon}), are orthogonal to the corresponding poloidal field $\Bv_{\St}$, defined by Equations (\ref{BScon}) and (\ref{GBS}).
Therefore, plotting equally spaced iso-contours of $\St$ and superimposing them on the corresponding distribution of
$
B_{\St}
\equiv
|
\Bv_{\St}
|
$
is also a natural way to visualize our tangential poloidal field at the surface.

As shown further on, this type of field visualization should be particularly useful for realistic magnetic configurations.
The surface sources,
$J_{r}$ and 
$
\nt
\bcd
\Btld
$, for the current-carrying part of the photospheric field, $\Btld$, in these configurations are usually represented by a myriad of small concentrations of different sizes, which are scattered semirandomly over the surface.
However, even for such complex sources, the corresponding iso-contours of $T$ and $\St$ reveal coherent field structures on length scales larger than the concentration sizes.

Although all coronal currents, regardless of their type and connectivity, contribute to our field
$
\Btld
=
\BvT
+
\BvSt
$
at
$
|\xv|=1_{+0}
$,
the main contribution to this field comes from currents flowing at low heights in the corona.
Therefore, the visualization of
$
\BvT
$
and
$
\BvSt
$
has to reveal the photospheric imprint of primarily these currents.
The contributions of the corresponding closure-current elements and elementary magnetic shells have only to enhance this imprint.
This is because they essentially play the same role as the image current in configurations with planar geometry, where coronal and mirrored current elements produce codirected contributions to the photospheric field $\Btld$.

As a rule, MFRs reside at low heights above and along segments of the polarity-inversion line (PIL).
The total axial current in such MFRs can often differ from zero or, in other words, be unneutralized on the length scale of the segment size.
Visualizing the fields
$
\BvT
$
and
$
\BvSt
$
around these segments then allows one (1) to establish this fact and (2) to determine the direction of the current.
Together with the iso-contours of $T$ and $\St$, this allows one to localize such MFRs in projection to the solar surface.
It should be emphasized that this important information is obtained by using only vector magnetic data without modeling the corresponding PECs themselves.

\subsubsection{
The Modeled PEC of the 2009 February 13 CME
	\label{s:Feb13}
}

Let us see how our decomposition method works in the case of a simple sigmoidal PEC model, which has previously been described as Solution 1 in \citet{Titov2021}.
It was found there that the core of this PEC contains an MFR embedded in a sheared magnetic arcade such that a substantial part of its electrical current is concentrated in layers at the central part of the PEC.
Panels (a) and (b) in Figure \ref{Feb13_Bph_dec} depict the top views of the corresponding current and magnetic field structures in the core.
Comparison of these structures with the results of the decomposition will help us to assess the potential of this method.

The photospheric distribution of $J_r$ obtained in the model is rather nontrivial. Panel (d) shows that this distribution comprises relatively large spots with low values of $J_{r}$ and narrow stripes with high values $J_{r}$ that stretch near and along the central part of the PIL.
In contrast, the TFFF, computed for this $J_r$ distribution by means of Equations (\ref{GSP}) and (\ref{Tcon}), shows a relatively simple pattern of equidistant iso-contours.
They clearly reveal two extrema within the MFR footprints, which are located at the periphery of the magnetic flux spots (see panels (d) and (e)).

Using Equations (\ref{BTcon}), (\ref{GBT}), (\ref{BScon}), and (\ref{GBS}), we computed the photospheric magnetic components $\BvT$ and $\BvSt$ whose directions in the region of interest are depicted by cyan and magenta arrows in panels (d) and (g), respectively.
They both match the direction of the modeled unneutralized MFR current that flows in our PEC from positive to negative magnetic polarity.
Indeed, first, as panel (d) shows, the directional field of $\BvT$ (cyan) forms a clockwise and counterclockwise vortex at positive and negative magnetic polarity, respectively.
Both vortices are centered around the footprints of the MFR.
Second, panel (g) shows that at the PIL segment, above which the MFR resides, the directional field of $\BvSt$ (magenta) is directed from negative to positive magnetic polarity.
These properties of $\BvT$ and $\BvSt$ evidently agree with what the curl right-hand rule would provide us, given the known location and direction of the MFR current.

Panels (f) and (i) present the grayscaled distributions of $B_T \equiv |\BvT|$ and $B_{\St} \equiv |\BvSt|$, on top of which the corresponding directional fields and iso-contours of $T$ and $\St$ are overlaid.
Based on the density of these iso-contours, $B_T$ and $B_{\St}$ are enhanced near the MFR in somewhat different ways.
In particular, the $B_T$ distribution is largely concentrated in the central part of the PIL and, to a lesser extent, outside the PIL by encircling the MFR footprints.
The $B_{\St}$ distribution forms two J-like hooks adjacent to the same central part of the PIL.
Both distributions outline a sigmoidal shape, which appears as a photospheric ``shadow'' of the corresponding MFR in the corona.

This sigmoidal shape is also visible in the $Q$-map of the region of interest presented in panel (c) as the distribution of $\slog Q$, which is essentially $\log_{10}$ of the squashing factor $Q$ taken with the sign of local $B_r$ \citep{Titov2011}.
The high-$Q$ lines generally mark the footprints of quasi-separatrix layers \citep[QSLs,][]{Priest1995, Demoulin1996, Demoulin1996a} formed by strongly divergent magnetic field lines.
QSLs serve as interfaces between magnetic flux systems with different types of field-line connectivity to the boundary.
The meaning of this, rather complex, $Q$-map for our PEC was previously considered in detail by \cite{Titov2021}.
Here we just point out that each magnetic polarity in this map contains a high-$Q$ line of J-like shape that wraps around one of the two extrema of the TFFF.
And, as explained above, these extrema are centered on the MFR footprints.

Thus, the distributions of both $Q$ and $T$ allow approximately the same location of the MFR footprints to be identified.
However, the $Q$-maps are calculated by using the magnetic field of PECs, the modeling of which
is technically a nontrivial and numerically expensive procedure.
In contrast, the calculation of $T$-maps requires only a relatively simple convolution of the photospheric $J_r$ distributions, which can be directly determined from vector magnetograms without modeling PECs themselves.
We can also anticipate that the calculations based on similar convolutions enable one to find the photospheric toroidal and poloidal fields, $\BvT$ and $\BvSt$, which, 
in combination with the iso-contours of $T$ and $\St$, provide valuable constraints for modeling MFR configurations.

\begin{figure*}[ht!]
\centering
\resizebox{0.99\textwidth}{!}{
\includegraphics{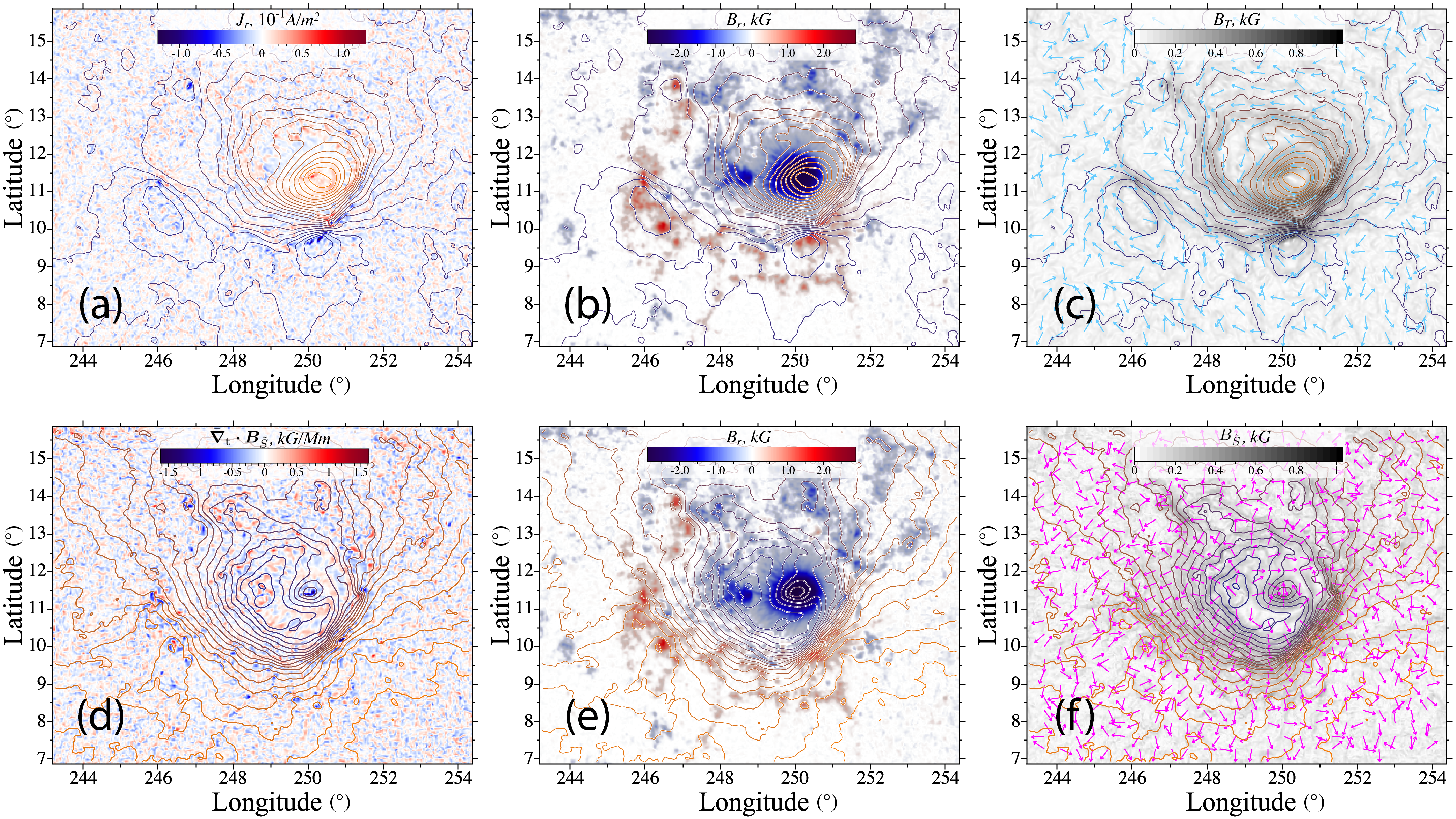}}
\caption{
Decomposition of the vector magnetic data for the AR 11305 at 9:36 UT
when the onset of the 2011 October 1 CME event occurred:
the same colors for the lines and arrows
as in panels (d)\textendash(i) of
Figure \ref{Feb13_Bph_dec} are used to show the directional fields of $\Bv_T$ (c) and $\BvSt$ (f) along with equally spaced iso-contours of $T$, superimposed on the corresponding distributions  of $J_r$ (a), $B_r$ (b), and $B_T$ (c),
and equally spaced iso-contours of $\St$, superimposed on the corresponding distributions of
$\bar{\bm{\nabla}}_{\mathrm{t}} \bcd \BvSt$ (d)
, $B_r$ (e), and $B_{\St}$ (f).
The blue-red distribution of $B_r$ is overlaid on the gray-shaded areas of $B_r\! \ge\! 100\:{\mathrm G}$ to bracket the PIL along which an MFR resides; the MFR splits into two parts with the footprints located at the extrema of $T$, one in positive polarity and two in negative one.
\label{Oct1_UNALT}}
\end{figure*}

\begin{figure*}[ht!]
\centering
\resizebox{0.99\textwidth}{!}{
\includegraphics{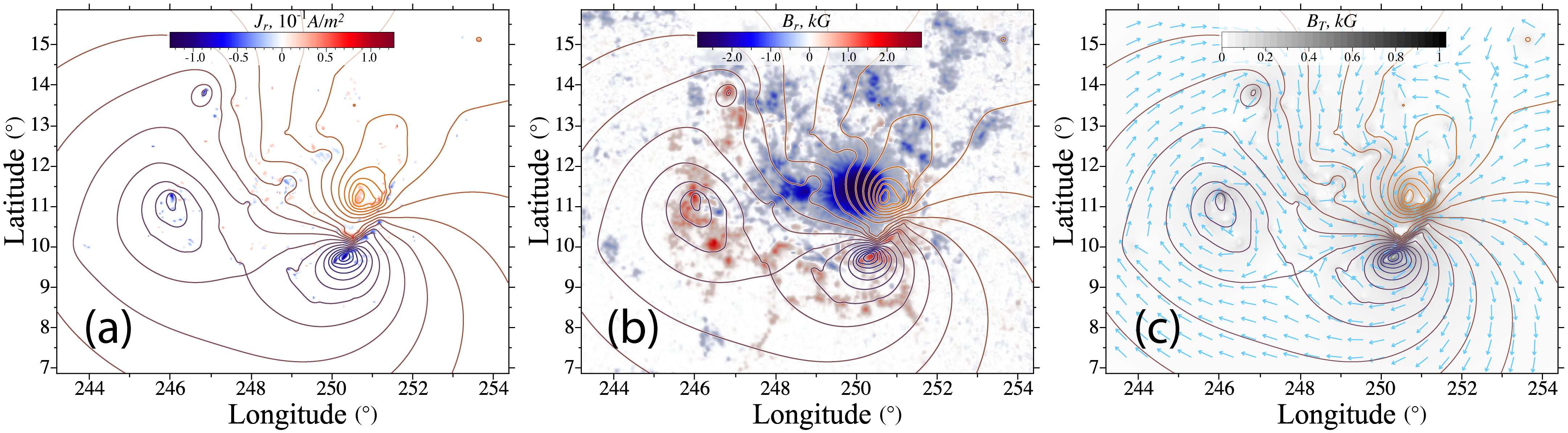}}
\caption{
A similar set of
{ the panels as in the first row of} Figure \ref{Oct1_UNALT} for the same vector magnetogram, but for the convolutions of the $J_r$ distribution (a) where weak-current spots with strengths below the estimated error of the measurements have been excluded.
\label{Oct1_JR-THRESH}}
\end{figure*}

\begin{figure*}[ht!]
\centering
\resizebox{0.99\textwidth}{!}{
\includegraphics{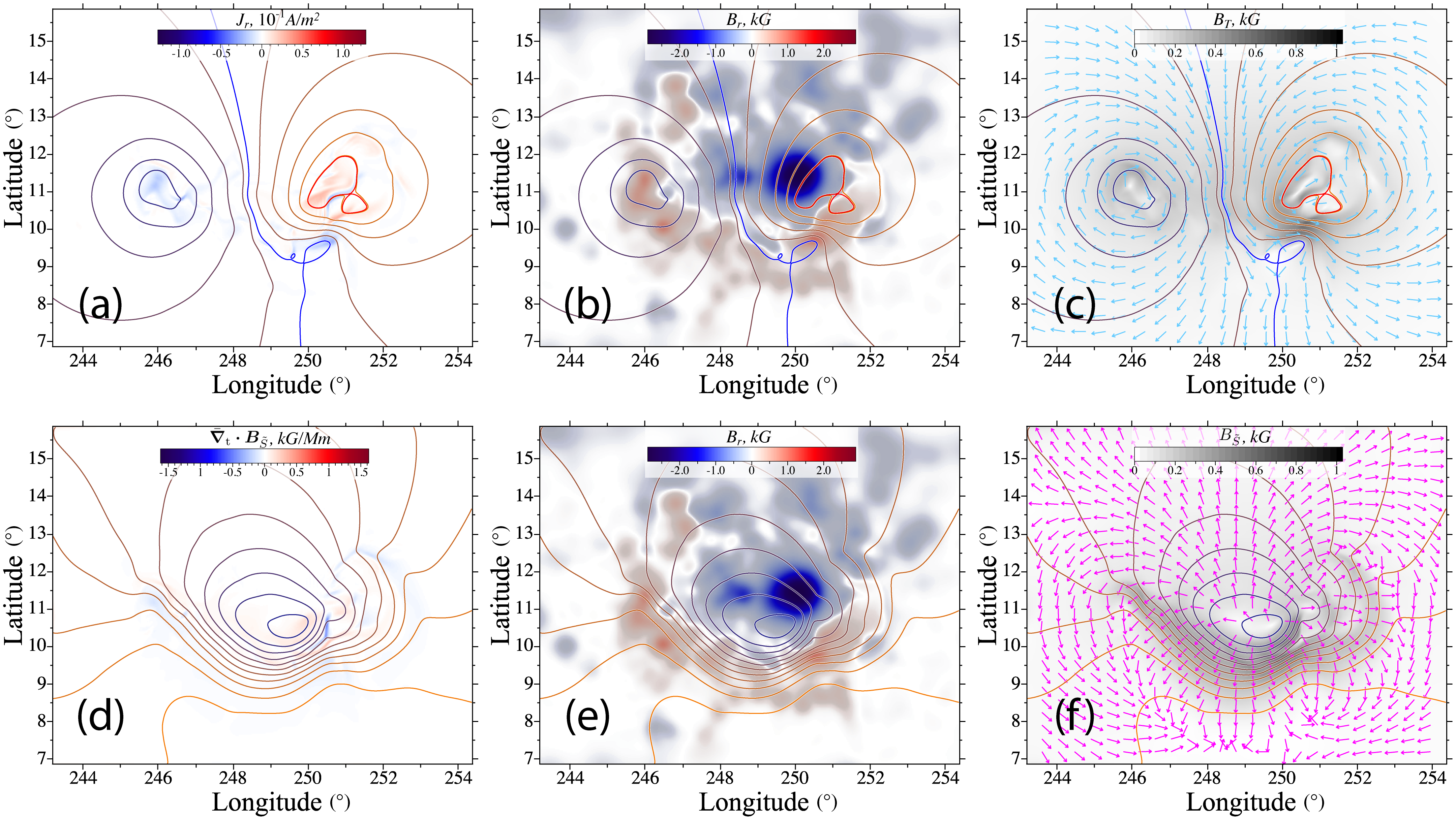}}
\caption{
A similar set of the panels as
in Figure \ref{Oct1_UNALT}, but for the decomposition of the photospheric magnetic field in the PEC of the 2011 October 1 CME event, which was modeled by $\beta\!=\!0$ MHD simulations.
The MFR was constructed in this PEC using \RBSL and helicity pumping methods under the constraints provided by the observed photospheric distribution of $B_r$ and the corresponding coronal extreme-ultraviolet (EUV) images, as described in V. S. Titov et al. (\citeyear{Titov2018}, \citeyear{Titov2021}, \citeyear{Titov2022}).
\label{Oct1_MODEL}}
\end{figure*}
\subsubsection{
The PEC at the Onset of the 2011 October 1 CME
	\label{ss:2011Oct1}
}

From a practical point of view, it is important to check how our decomposition technique works with real vector magnetograms.
For this purpose, let us apply it to the HMI SHARP cea vector magnetic data for the AR 11305 obtained approximately at 9:36 UT when the 2011 October 1 CME event started.
In the following, we present the resulting decomposition for two different maps of $J_r$.

The first of these maps is derived by simply taking appropriate finite differences of the tangential components of the magnetogram.
The decomposition based on this map is illustrated in Figure \ref{Oct1_UNALT},
where panel (a) demonstrates that the distribution of $J_{r}$ is fragmented in numerous negative and positive spots of different strengths and sizes.
The current spots of small strengths and sizes appear to be randomly distributed throughout the AR.
In contrast, a signification fraction of current spots of larger strengths and sizes is aggregated in two unipolar necklace-like structures of opposite signs.
As seen in the region near the largest spot of negative flux in panels (a) and (b), these ``necklaces'' stretch along the PIL with a shift to each other, each on its own side of the PIL.

Despite such a complexity of the $J_r$-map, the corresponding equally spaced iso-contours of $T$ form a remarkably simple and coherent pattern, which clearly reveals the presence of three extrema of $T$.
One of them is a maximum located at the largest spot of negative flux, while the other two are minima located at two separate aggregations of small positive flux spots (see panel (b)).
The randomness mentioned above of weak-current spots manifests itself only in a noticeable jaggedness of the iso-contour lines.
Apparently, this property of $T$ is due to the averaging of counteracting contributions from different current spots in the AR.
Indeed, the contributions to $T$ from two neighboring current spots of similar strengths and sizes but of opposite signs have to partially cancel each other out in the convolution defined by Equations (\ref{GSP}) and (\ref{Tcon}).

Panel (c) complements this information by showing the gray-shaded distribution of $B_{T}$
overlaid with the corresponding directional field of $\BvT$ (cyan) and the iso-contours of $T$.
One can see from this panel that the toroidal field $\Bv_T$ is strongly sheared at the PIL and concentrated at the current-spot ``necklaces''.
This fact is also confirmed by the overlaid iso-contours, which tend to align with and condense in the ``necklaces.''

An additional complementary information on our configuration is obtained by visualizing the corresponding poloidal field $\BvSt$, the key constituent of which are the equally spaced iso-contours of $\St$.
Taking the appropriate finite differences of $\Btld=\Bv-\Bpot$ at the boundary, we first determine its surface divergence, which is equal to $\nt \bcd \BvSt$.
Then, taking the convolutions defined by Equations (\ref{GSP}) and (\ref{Scon})\textendash(\ref{GBS}), we calculate the desired $\St$ and $\BvSt$.
Panels (d)\textendash(f) in Figure \ref{Oct1_UNALT} depict the obtained results.

The distribution of $\bar{\bm{\nabla}}_{\mathrm{t}} \bcd \BvSt \equiv \nt \bcd \BvSt / R_{\sun}$ shown in panel (d) is as fragmented and irregular as the distribution of $J_{r}$  for this configuration.
Nevertheless, the iso-contours of $\St$ form a nice coherent pattern.
Similarly to the relationship between $J_{r}$ and $T$, the irregularity of the $\bar{\bm{\nabla}}_{\mathrm{t}} \bcd \BvSt$ distribution manifests itself only in a noticeable jaggedness of the iso-contour lines of $\St$.

Panels (e)\textendash(f) show that the field $\BvSt$ is localized in the PIL and particularly at the ``necklaces''.
However, in contrast to $\BvT$, it practically has no shear by traversing the iso-contours of $\St$ perpendicularly out from the largest spot of negative flux.
This mutual orientation of the field $\BvSt$ and iso-contours of $\St$ passing through the ``necklaces'' remains qualitatively the same when going eastward along these iso-contours from one minimum of $T$ to the other.

Similarly to the case considered in Section \ref{s:Feb13},
the described decomposition of the photospheric field can be interpreted as follows.
The configuration in the study contains two current channels, which are presumably MFRs that jointly start at the maximum of $T$ (counterclockwise vortex) and separately end at one of the two minima of $T$ (clockwise vortices).
The direction of these currents qualitatively matches what the curl right-hand rule requires for the directions of both the vortex circulations and the field $\BvSt$ at those iso-contours of $T$ that pass through the current-spot ``necklaces.''
As shown in the following, this interpretation also compares well with a numerical PEC model that we have developed earlier by using only the photospheric distribution of $B_{r}$ from the available magnetic data.

However, before doing this comparison, let us assess how sensitive our results are to the errors of magnetic field measurements.
For this purpose, we have made a similar decomposition of the same vector magnetic data using a modified $J_{r}$-map, which is obtained by cleaning up the previous one from the current values with a large uncertainty of the measurement.
The uncertainty in the current for each pixel is propagated from the provided tangential magnetic field error data.
Errors due to tangential field disambiguation are recorded by the conf\_disambig parameter \citep{Hoeksema2014}. The pixels are kept in the cleaned map if they satisfy the following conditions:
(1) their current is greater than 1.0 times the corresponding error, and (2) they
have high confidence in the disambiguation algorithm.
This cleaning procedure essentially removed most of the randomly distributed weak-current spots from the data, while keeping the strong-current spots.
In particular, the two necklace-shaped structures mentioned above have been preserved in the resulting $J_{r}$-map (panel (a) in Figure \ref{Oct1_JR-THRESH}).

Compared to the previous $J_{r}$-map, the new one provides the distribution of $T$ with much smoother iso-contours, because incoherent weak-current spots no longer contribute to the corresponding convolution of $J_{r}$.
However, the equally spaced iso-contours of $T$ in panel (b) demonstrate similar patterns: They reveal again two clockwise vortices and one counterclockwise vortex at approximately the same locations as before.
However, these patterns now correspond to a different set of $T$ values, since many weak-current spots have not been included in the convolution.
The exclusion of positive weak-current spots, those that are grouped within the largest negative flux spot (see panels (a) and (b) in Figure \ref{Oct1_UNALT}), causes only a small shift of the counterclockwise vortex to the border of this flux spot (cf. panels (b) in Figures \ref{Oct1_UNALT} and \ref{Oct1_JR-THRESH}).

We see that despite the significant differences between the original and cleaned $J_{r}$-maps, both decompositions reveal the possible presence and location of the two MFRs in the PEC under study.
Therefore, such an outcome of our decomposition procedure appears to be rather robust to errors of measurement of the photospheric magnetic field.
This is because, by construction, the distribution of $T$ is relatively insensitive to incoherent, even if multiple, small spots of the input $J_{r}$ distributions.

An additional evidence of the latter we find by comparing the above decomposition results with those that refer to our $\beta=0$ MHD modeling of the 2011 October 1 CME event.
The results of the field decomposition for this model are shown in
Figure \ref{Oct1_MODEL}, which presents panels similar to those of Figures \ref{Oct1_UNALT} for the moment when the modeled MFR starts to erupt.

This model was constructed long before the development of our decomposition technique.
The first steps of this model construction, using our \RBSL{} method to build and optimize an MFR in the PEC under study, were described in V. S. Titov et al. (\citeyear{Titov2018}, \citeyear{Titov2021}).
Later, the constructed PEC was energized toward an eruption by applying our helicity pumping method \citep{Titov2022}.

Note that the modeled PEC was constrained by using only the observed photospheric $B_{r}$  distribution and the corresponding extreme-UV (EUV) images of the AR.
These images were used, in particular, to identify possible locations of the MFR footprints, which were needed, in turn, to construct our initial \RBSL{} MFR.
By comparing Figures \ref{Oct1_UNALT}\textendash\ref{Oct1_MODEL} one can see that the extrema
and corresponding vortices of $T$ derived for our modeled and observed vector magnetic data match well enough at the footprints of the MFR.
A similar conclusion about the poloidal field $\BvSt$ can be drawn for the region where the MFR is located by comparing panels (d)\textendash(f) in Figures \ref{Oct1_UNALT} and \ref{Oct1_MODEL}.

Moreover, during the MHD relaxation of our initial approximate equilibrium, our initially single \RBSL{} MFR split to produce another MFR of shorter length.
To the end of the relaxation, a distinct footprint of the new MFR has been formed in the positive magnetic polarity, while its footprint in the negative polarity remained unified with the footprint of the initial MFR.
The locations of both these footprints are consistent with what our decomposition of the observed vector magnetogram predicts.
It is remarkable that such a good match occurred in spite of significant differences in the corresponding $J_{r}$-maps.

\section{Summary and conclusions
	\label{s:sum}}

When modeling PEC equilibria under constraints provided by observed magnetic data, it is convenient to treat the current-carrying and potential parts of the PEC separately, at least as far as the contributions of these parts to the radial magnetic field component, $B_{r}$, at the boundary are concerned.
We have successfully used this separation in our previously proposed \RBSL{} method for constructing
MFR equilibria (Titov et al. \citeyear{Titov2018}, \citeyear{Titov2021}).
However, this was done in a restricted form that only allows one to efficiently construct MFRs with lengths smaller than the solar radius.
The present work overcomes this limitation in our previous version of the \RBSL{} method.
In addition, the new approach presented in this paper could be used to extend other methods,
particularly those that explicitly use \BSL{} to construct PEC equilibria \citep[e.g.,][]{Wheatland2004, Wheatland2007, Gilchrist2014}.
To this end, we have derived the \MBSL{} for a coronal current that can be concentrated at a given path or distributed in the volume (see Equations (\ref{mmBSL1r}) and (\ref{mmBSLc}), respectively). 
By definition, \MBSL{} determines the magnetic field of this current under the condition that only a tangential field component is produced on the photospheric surface.
We achieved this by introducing for every \BSL{} current element, irrespective of whether it belongs to coronal currents or to the subsurface closure current, an auxiliary fictitious source of a potential magnetic field, given by Equations (\ref{dBCs}) and (\ref{dBC}). 
This is done in such a way that the radial components of the current element and the associated fictitious source at the surface compensate for each other.
These elementary sources of the compensating field are represented by magnetized triangular shells, one vertex of which is located at the center of the Sun and two others below the surface at an infinitesimal distance from each other.

Using this \MBSL, we derived an elegant expression, given by Equation (\ref{AmmBSL1r}), which provides the magnetogram-matching vector potential of a line current of arbitrary shape.
The regularized version of this expression, given by Equation (\ref{AmmRBSL}), substantially improves our \RBSL{} method, in particular, the iterative optimization procedure for finding an MFR shape with minimized Lorentz forces \citep[see][]{Titov2021}. The modified procedure now allows one to keep the same background potential field throughout all iterations of the optimization, regardless of the length of the PEC (filament channel) to be modeled.
 
Applying our approach solely to the subphotospheric closure current, we then derived
that the field it produces in the corona and on the surface is purely toroidal. 
This field has no radial component and is expressed in terms of the convolution of the photospheric radial current density, $J_{r}$, and the corresponding source function (see the last term in Equation (\ref{mmBSL2}) and the corresponding vector potential represented by Equations (\ref{ATv})\textendash(\ref{GT1})).
We demonstrated that elementary contributions to this convolution originate from the radial edge currents of our elementary magnetic shells.
It is of particular importance that this toroidal field does not depend on the shape of the closure currents, which implies that these currents manifest themselves in the corona only by means of the surface $J_{r}$ distributions. 
However, we have shown that this field is approximately one-half of the total toroidal field $\BvT$ on the surface.
The remaining half of $\BvT$ is generated by the coronal currents that, together with the subphotospheric closure currents, form full circuits in space.

Based on these results, 
we have developed a new method for decomposing an observed
photospheric magnetic field $\Bv$ into the following three parts: (1) 
the potential field $\Bpot$ calculated from the observed $B_{r}$, (2) the total toroidal field $\BvT$ calculated from the observed $J_{r}$, and (3) the tangential poloidal field $\BvSt \equiv \Bv -\Bpot - \BvT $.
Part (1) is generated by the subphotospheric currents that circulate within the solar interior without reaching the surface.
Part (2) is generated by the currents that pass through the solar surface into the corona.
Part (3) is associated with all coronal currents, regardless of whether they reach the surface or not.
It is generated by these and subphotospheric closure currents together with all of our fictitious sources.
The latter are represented by magnetic shells that are set up on the ruled surfaces, which are
formed by a continuum of straight lines connecting the center of the Sun with the points of the corresponding closure-current paths or of the inversion images of the coronal-current paths. 
Part (3) of this decomposition can independently be obtained from the surface divergence of $\Bv - \Bpot$, which gives the advantage to express $\BvSt$ as a surface gradient of the spheroidal potential $\St$.

Part (2) in our field decomposition is the same as in the one recently proposed by \citet{Schuck2022}.
However, their other two parts differ very much from ours: These are the potential poloidal fields $\BPlt$ and $\BPgt$, which are generated separately by subphotospheric and coronal currents, respectively, at the upper and lower sides of the boundary.
Nonetheless, after reassigning $\BPgt$ to the upper side of the surface, the equality
$\BPlt + \BPgt \equiv \Bpot + \BvSt$ 
must hold, since 
all parts of the decomposed field in our method are defined at this level.
In Appendix \ref{s:relation}, we demonstrate that this is correct for a simple 2.5D magnetic configuration with a continuous current density distribution across the boundary.
We also argue that this should
be true for any 3D configuration under the same continuity condition of the current density.

The effect of coronal currents on photospheric $B_{r}$ is eliminated in our approach by the compensating magnetic field, which is generated in the solar interior by a fictitious closed-current system of magnetic shells.
This makes it possible to relate an observed photospheric $B_{r}$ completely to subphotospheric currents that circulate entirely within the interior.
Thus, the total magnetic field in the corona is represented then as a sum of the potential field defined by $B_{r}$ and the field produced by coronal currents.

In addition, our decomposition enables one to see how the photospheric field of coronal currents would look if the solar globe were an ideal rigid conductor that shields the interior from the magnetic field generated by coronal currents.
In other words, it incorporates, in an idealized form, the response of the dense photospheric and subphotospheric layers to fast variations of coronal currents, such as those occurring during solar eruptions.
Therefore, our decomposition should be useful for the analysis of such variations.
For example, it makes it possible to derive, from a sequence of vector magnetic data, the surface currents induced during eruptions and the corresponding Lorentz forces.

We demonstrated that our field decomposition allows one to reveal (1) the location of an MFR or, more generally, a coronal-current channel, in projection to the photospheric surface,  particularly its footprint locations, and (2) the direction of an unneutralized MFR current before modeling the corresponding PEC.
Moreover, the detection of additional current channel footprints and the poloidal field pattern in the region of interest, as for the case described in Section \ref{ss:2011Oct1}, can yield further important insights about the corresponding coronal magnetic fields.
This provides valuable constraints for PEC modeling, as well as important information for the analysis of erupting and post-eruptive configurations and the interpretation of the corresponding observations taken in, e.g., EUV wavelengths.
Regarding the determination of the projected location of an MFR on a given vector magnetogram, it has yet to be seen whether the poloidal parts of our decomposition and the decomposition of \citet{Schuck2022} provide similar results in this respect.


\begin{acknowledgments}

We are very grateful to Mark Linton and Peter Schuck for the valuable discussions that helped us to improve this paper.
This research was supported by NASA grants 80NSSC20K1317, 80NSSC22K1021, 80NSSC20K1274, 80NSSC22K0893, 80NSSC19K0858, and 80NSSC24K1108, 
by NSF grants AGS-1923377 and ICER1854790, and by the PSP WISPR contract NNG11EK11I to N.R.L. (under subcontract N00173-24-C-0004 to PSI). Computational resources were provided by NSF's XSEDE and NASA's NAS.

\end{acknowledgments}
\appendix
\section{
Compensating Magnetic Field}
\label{s:dBC}

To derive the compensating magnetic field, let us choose our global Cartesian system of coordinates such that its $z$-axis is directed along $\Rv$, which means that
\begin{eqnarray}
\Rv
=
{\mathcal R}
\left(
0
,
0
,
1
\right)
\end{eqnarray}
and
\begin{eqnarray}
\xv
=
|\xv|
\left(
\cos\phi
\sin\theta
,
\,
\sin\phi
\sin\theta
,
\,
\cos\theta
\right)
\,
,
	\label{xv}
\end{eqnarray}
where $\theta$ and $\phi$ are latitude and longitude, respectively, of the spherical system of coordinates whose center is the same as for the Cartesian system.

Then, for the displacement vector
\begin{eqnarray}
{\mathrm d}
\rv
=
-
{\mathrm d}
\Rv
=
-
\left(
{\mathrm d}
{\mathcal X}
,
{\mathrm d}
{\mathcal Y}
,
{\mathrm d}
{\mathcal Z}
\right)
\,
,
	\label{drv}
\end{eqnarray}
we have the following negative radial component of the elementary \BSL{} field:
\begin{eqnarray}
-
\hat{\xv}
\bcd
{\mathrm d}
\bm{B}_{I}
\bigr|
_{|\xv|=1}
=
-
\left.
\frac
{
\hat{\xv}
\bcd
\rv
\bmt
{\mathrm d}
\rv
}
{
r^3
}
\right|
_{|\xv|=1}
=
-
\frac
{
{\mathcal R}
\sin\theta
\left(
\sin\phi
\,
{\mathrm d}
{\mathcal X}
-
\cos\phi
\,
{\mathrm d}
{\mathcal Y}
\right)
}
{
r_1^3
}
\,
,
	\label{x_dBI_bc}
\end{eqnarray}
where
\begin{eqnarray}
r_1
=
\sqrt
{
1
-
2
\,
{\mathcal R}
\cos\theta
+
{\mathcal R}^2
}
\,
.
	\label{r1}
\end{eqnarray}

\subsection{Subphotospheric Path $\Cs$
	\label{A1}}

We are looking for the compensating potential magnetic field such that 
\begin{eqnarray}
{\mathrm d}
\bm{B}_{\Cs}
=
-
{\bm \nabla}_{\xv}
\left(
{\mathrm d}
F_{\Cs}
\right)
\,
,
	\label{dBCs_gen}
\end{eqnarray}
where the potential
${\mathrm d}
F_{\Cs}
$
is a regular harmonic function at $|\xv|>1$ that satisfies the Laplace equation
\begin{eqnarray}
{\bm \nabla}_{\xv}^{2}
\left(
{\mathrm d}
F_{\Cs}
\right)
=
0
\,
,
	\label{LEq}
\end{eqnarray}
and the following boundary condition:
\begin{eqnarray}
\left.
-
\frac
{\partial\ }
{
\partial
|\xv|
}
{\mathrm d}
F_{\Cs}
\right|
_{|\xv|=1}
=
-
\hat{\xv}
\bcd
{\mathrm d}
\bm{B}_{I}
\bigr|
_{|\xv|=1}
\,
.
	\label{dFCs_bc}
\end{eqnarray}
Thus, we obtain for
$
{\mathrm d}
F_{\Cs}
$
the external Neumann problem with the spherical boundary $|\xv|=1$.
Instead of applying a standard method for solving this problem, let us use a more heuristic approach that exploits a relatively simple form of the boundary condition defined by Equations (\ref{x_dBI_bc}) and (\ref{dFCs_bc}).

Note first that this condition suggests that the following relationship
\begin{eqnarray}
\frac
{\partial\ }
{
\partial
|\xv|
}
{\mathrm d}
F_{\Cs}
&=&
\frac
{
{\mathcal R}
\sin\theta
\left(
\sin\phi
\,
{\mathrm d}
{\mathcal X}
-
\cos\phi
\,
{\mathrm d}
{\mathcal Y}
\right)
}
{
r^3
}
\,
,
	\label{assum}
\\
r
&=&
\sqrt
{
|\xv|^2
-
2
\,
|\xv|
{\mathcal R}
\cos\theta
+
{\mathcal R}^2
}
=
\left|
\xv
-
\Rv
\right|
\,
,
\end{eqnarray}
possibly holds for $|\xv|$  other than $|\xv|=1$ as well.
To verify this strong assumption, let us integrate Equation (\ref{assum}) over $|\xv|$ to obtain
\begin{eqnarray}
{\mathrm d}
F_{\Cs}
=
\frac
{
{\mathcal R}
\cos\theta
-
|\xv|
}
{
r
\,
{\mathcal R}
\sin\theta
}
\left(
\cos\phi
\,
{\mathrm d}
{\mathcal Y}
-
\sin\phi
\,
{\mathrm d}
{\mathcal X}
\right)
+
\delta(\theta,\phi)
\,
,
	\label{dFCs_try}
\end{eqnarray}
where $\delta(\theta,\phi)$ is generally an arbitrary function, which can also depends on ${\mathrm d}\Rv$ and $R$ as on parameters.
One can prove by direct substitution that the first term of Equation (\ref{dFCs_try}) is a solution of Equation (\ref{LEq}).
However, this heuristic solution of serendipity is singular at $\sin\theta=0$, which corresponds to a nonlocal singularity extended throughout the whole space.
The latter property is not acceptable for us, because our solution must be regular at $|\xv|>1$.

Fortunately, this issue can be resolved by using, in Equation (\ref{dFCs_try}), the second term
$
\delta(\theta,\phi)
$, which then also is to be an irregular harmonic function whose singularity, however, should eliminate the singularity of the first term in the domain of interest, namely, at $|\xv|>1$.
The desired solution
$
\delta(\theta,\phi)
$
of the Laplace equation is easily found, as it does not depend on $|\xv|$.
The result reads as follows:
\begin{eqnarray}
\delta(\theta,\phi)
=
\frac
{1}
{
{\mathcal R}
\sin\theta
}
\left(
\cos\phi
\,
{\mathrm d}
{\mathcal Y}
-
\sin\phi
\,
{\mathrm d}
{\mathcal X}
\right)
\,
,
\end{eqnarray}
so that Equation (\ref{dFCs_try}), after some algebraic calculations, becomes
\begin{eqnarray}
{\mathrm d}
F_{\Cs}
=
\frac
{
{\mathcal R}
\sin\theta
\left(
\cos\phi
\,
{\mathrm d}
{\mathcal Y}
-
\sin\phi
\,
{\mathrm d}
{\mathcal X}
\right)
}
{
r
\left(
|\xv|
-
{\mathcal R}
\cos\theta
+
r
\right)
}
\,
.
	\label{dFCs_co}
\end{eqnarray}

The obtained potential
$
{\mathrm d}
F_{\Cs}
$
tends to infinity if the denominator in Equation (\ref{dFCs_co}) vanishes.
This occurs at 
$
\theta=0
$ and 
$
0 
\le 
|\xv| 
\le 
{\mathcal R}
$
when 
$
|\xv|
-
R
+
r
=0
$.
However, the resulting singularity is acceptable for ${\mathcal R}<1$, since it is located within the photospheric surface $|\xv|=1$, and therefore, our solution is regular in the corona $|\xv|>1$, as required. 

Since we are going to use our Equation (\ref{dFCs_co}) for an arbitrary  element of the current path, it is useful to rewrite it in the vector form as follows:
\begin{eqnarray}
\left[
\frac
{\mu I}
{4\pi}
\right]
\qquad
{\mathrm d}
F_{\Cs}
=
\frac
{
\rv
\bmt
\xv
\bcd
{\mathrm d}
\rv
}
{
r
\left(
\xv
\bcd
\rv
+
|\xv|
\,
r
\right)
}
\,
.
	\label{dFCs_vf}
\end{eqnarray}
Then, using this expression and Equation (\ref{dBCs_gen}) the compensating field required,
$
{\mathrm d}
\bm{B}
_{\Cs}
$,
is obtained; it is presented above by Equation (\ref{dBCs}).

To understand the physical meaning of Equation (\ref{dFCs_vf}), 
let us substitute the relationships $\xv=\Rv+\rv$ and $ {\mathrm d} \rv = - {\mathrm d} \Rv $ into it
to obtain
\begin{eqnarray}
\left[
\frac
{\mu I}
{4\pi}
\right]
\qquad
{\mathrm d}
F_{\Cs}
=
\frac
{
{\mathrm d}
\Rv
\bmt
\Rv
\bcd
\rv
}
{
r^2
\,
\left(
r
+
\Rv
\bcd
\hat{\rv}
+
\left|
\rv
+
\Rv
\right| 
\right)
}
\,
.
	\label{dFCs_vfsc}
\end{eqnarray}
In this source-centric form, the asymptotic behavior of
$
{\mathrm d}
F_{\Cs}
$
becomes obvious: the leading term of its expansion by $r \gg {\mathcal R}$ is
\begin{eqnarray}
{\mathrm d}
F_{\Cs}
\sim
\frac
{
{\mathrm d}
\Rv
\bmt
\Rv
\bcd
\rv
}
{
2
\,
r^3
}
\,
,
	\label{dFCs_asym}
\end{eqnarray}
which is nothing else as the potential of the magnetic moment
$
\frac{1}{2}
{\mathrm d}
\Rv
\bmt
\Rv
$
normalized to
$
\left.
{
\mu
I
R^{2}_{\sun}
}
\right/
{4\pi}
$.
Further analysis of the behavior of Equation (\ref{dFCs_vfsc}) near the singularity mentioned above suggests that this moment is linearly distributed along the vector $\Rv$ from ${\mathbf 0}$ to
$
{\mathrm d}
\Rv
\bmt
\Rv
$.
For verification of this assumption, it is convenient to use the Cartesian system of coordinates whose origin is located at the point $\Rv$ and the $x$-, $y$- and $z$-axes are parallel to $\Rv$,
$
\left(
{\mathrm d}
\Rv
\bmt
\Rv
\right)
\bmt
\Rv
$,
and
$
{\mathrm d}
\Rv
\bmt
\Rv
$, 
respectively.
The linear superposition of the potentials generated by the linear distribution of the magnetic moment is given in these coordinates by the following integral:
\begin{eqnarray}
\frac
{
|
{\mathrm d}
\Rv
\bmt
\Rv
|
z
}
{
{\mathcal R}^2
}
\int
^{0}
_{-{\mathcal R}}
\frac
{
\left(
\xi
+
{\mathcal R}
\right)
{\mathrm d}
\xi
}
{
\left(
\xi^2
-
2
x
\xi
+
r^2
\right)
^{3/2}
}
=
\frac
{
|
{\mathrm d}
\Rv
\bmt
\Rv
|
\,
z
}
{
r
\left(
r^2
+
{\mathcal R}
x
+
r
\sqrt
{
r^2
+
2
{\mathcal R}
x
+
{\mathcal R}^2
}
\right)
}
\,
,
	\label{dFCs_scsc}
\end{eqnarray}
where
\begin{eqnarray}
r
=
\sqrt
{
x^2
+
y^2
+
z^2
}
\,
.
\
\end{eqnarray}
Rewriting now the result of this integration in terms of vectors, we arrive at the expression given by Equation (\ref{dFCs_vfsc}) and therefore validate our guess about its origin.

However, the derived solution admits another, more instructive, and deeper interpretation than the present.
Note, first, that the considered magnetic singularity refers, strictly speaking, not to the vector $\Rv$ itself, but rather to the infinitesimal triangle spanned by the vectors $\Rv$ and
$
{\mathrm d}
\Rv
$
as shown in Figure \ref{f:path}.
One can imagine that the area of this triangle is swept out by the radius vector $\Rv$ as a result of an infinitesimal displacement
$
{\mathrm d}
\Rv
$ of its head along the path $\Cs$.
The swept area equals
$
\frac{1}{2}
\left|
{\mathrm d}
\Rv
\bmt
\Rv
\right|
$,
which is exactly the dimensionless strength of the magnetic moment that we found above to be linearly distributed along $\Rv$ or, in view of the latter remark, over the infinitesimal triangle.

To relate the line and surface densities of the magnetic moment, let us consider similar triangles obtained from the indicated one via its homothety with respect to the solar center O.
With the homothetic coefficient $k$ running from 1 to 0, the area of these triangles and its increment scale as $k^2$ and $2k$, respectively.
This implies that the linear distribution of the line density of the magnetic moment along $\Rv$ is actually due to a uniform distribution of the magnetic moment over our triangle spanned by the vectors $\Rv$ and
$
{\mathrm d}
\Rv
$.
Normalized to $\mu I /4\pi$, the corresponding surface density of the magnetic moment in this triangle is equal to just unity.

In other words,  our infinitesimal triangle is magnetically polarized and is known in textbooks as a magnetized or magnetic shell \citep[see, e.g,][]{Stratton1941}.
Therefore, up to a coefficient proportional to the electric current associated with the magnetic shell,
its magnetic potential must be equal to the solid angle from the observation point subtended by this triangle.
From the above consideration of the homothety in the triangle,  it follows that the indicated solid angle is
\begin{eqnarray}
{\mathrm d}
\Omega
=
\left(
{\mathrm d}
\Rv
{\bm \times}
\Rv
\bcd
\xv
\right)
\int^{1}_{0}
\frac
{
k
\,
{\mathrm d}
k
}
{
\left|
\xv
-
k
\Rv
\right|
^{3}
}
\,
.
	\label{dOmg}
\end{eqnarray}
The integral here can be taken exactly and transformed into the expression given by Equation (\ref{dFCs_vfsc}), which straightforwardly confirms our physical interpretation of the potential
$
{\mathrm d}
F_{\Cs}
$

The integration of
$
{\mathrm d}
F_{\Cs}
$
itself or the corresponding
$
{\mathrm d}
\bm{B}_{\Cs}
$,
given by Equation (\ref{dBCs}),
along the path $\Cs$ provides the total potential
$
F_{\Cs}
$
or the compensating field
$
\bm{B}_{\Cs}
$,
respectively.
Thus, they are generated by the magnetic shell that geometrically is a ruled surface $\Ss$ swept out by the vector $\Rv$ when its head slides from the foot point $\Rv_{2}$ to $\Rv_{1}$ along the path $\Cs$, i.e. the directrix of $\Ss$  (see Figure \ref{f:ScSs}).
The resulting surface $\Ss$ is a curvilinear triangle with two straight sides and one curved represented by the vectors $\Rv_{2}$ and $\Rv_{1}$, and the path $\Cs$, respectively.

As shown above, the surface density of the magnetic moment is a unit vector field, say $\mvh$, normal to our infinitesimal triangles, and so to $\Ss$ itself.
According to the characteristic property of magnetic shells \citep[][]{Stratton1941}, infinitesimal currents, circulating within $\Ss$ to create the field $\mvh$, compensate one another throughout $\Ss$ except for its edges or sides.
Currents flowing along the edges and the original circuit $\Cc \cup \Cs$ have the same values and directions of circulation.
As a result of that, they are counter-directed at the path $\Cs$ and cancel each other out.
Thus, only the magnetic shell currents flowing along the straight sides of $\Ss$ are responsible for generating the field $\bm{B}_{I\Cs}$ represented  by Equations (\ref{Thev}) and (\ref{BICs}),
while the contribution to this field by the closure current flowing along the path $\Cs$ is fully compensated.

The latter provides, first, an elegant physics-based proof of the fact that the field $\bm{B}_{I\Cs}$ does not depend on the shape of the path $\Cs$ (see Section \ref{ss1:mmBSl}).
Second, it provides an alternative and more transparent way to explicitly determine $\bm{B}_{I\Cs}$: it can now be easily obtained by a simple integration of the elementary \BSL{} field, given by Equation (\ref{dBI}), along the straight sides of $\Ss$.
Finally, establishing the relation between $\bm{B}_{I\Cs}$ and the magnetic shell, we gain a deep insight into the nature of the toroidal magnetic field.
Indeed, Equations (\ref{rotBICs})\textendash(\ref{BICs2}) reveal that $\bm{B}_{I\Cs}$ is a toroidal field whose current density is the potential field generated by two point sources of opposite signs.
We now see that these sources are located at the upper end points of the straight sides of $\Ss$, which are also the foot points $\Rv_1$ and $\Rv_2$, as shown before.
This coronal current, therefore, provides closure to the edge currents on the straight sides of the magnetic shell $\Ss$.

\subsection{Coronal Path $\Cc$
	\label{A2}}

The compensating field
$
{\mathrm d}
\bm{B}_{\Cc}
$
and its harmonic potential
$
{\mathrm d}
F_{\Cc}
$
for the coronal path $\Cc$ is determined similarly to that for the path $\Cs$, except for the following point.
The singularity of
$
{\mathrm d}
F_{\Cc}
$
cannot be distributed over the entire vector $\Rv$, since ${\mathcal R} \ge 1$ and so this singularity would extend into the corona making the corresponding
$
{\mathrm d}
F_{\Cc}
$
unacceptable.
Instead, it is natural in this case to try and use for carrying the singularity the vector $\Rv_{*}$, which is the image of $\Rv$ due to its inversion given by Equation (\ref{Rvs}).
By definition, its length ${\mathcal R}_{*}=1/{\mathcal R} \le 1$ and hence the singularity would be fully contained within the sphere $|\xv|=1$, as required.

Based on this consideration, we substitute ${\mathcal R}_{*}$ for ${\mathcal R}$ in Equation (\ref{dFCs_co}) and, for the reason explained below, multiply it additionally on ${\mathcal R}_{*}$ to obtain
\begin{eqnarray}
{\mathrm d}
F_{\Cc}
=
\frac
{
{\mathcal R}_{*}^{2}
\sin\theta
\left(
\cos\phi
\,
{\mathrm d}
{\mathcal Y}
-
\sin\phi
\,
{\mathrm d}
{\mathcal X}
\right)
}
{
r_{*}
\left(
|\xv|
-
{\mathcal R}_{*}
\cos\theta
+
r_{*}
\right)
}
\,
,
	\label{dFC_co}
\end{eqnarray}
where
\begin{eqnarray}
r_{*}
=
\sqrt
{
|\xv|^2
-
2
\,
|\xv|
{\mathcal R}_{*}
\cos\theta
+
{\mathcal R}_{*}^2
}
=
\left|
\xv
-
\Rv_{*}
\right|
\,
.
\end{eqnarray}

In vector form, similar to Equation (\ref{dFCs_vf}), this expression is written as follows:
\begin{eqnarray}
\left[
\frac
{\mu I}
{4\pi}
\right]
\qquad
{\mathrm d}
F_{\Cc}
=
\frac
{1}
{{\mathcal R}}
\frac
{
\rv_{*}
\bmt
\xv
\bcd
{\mathrm d}
\rv
}
{
r_{*}
\left(
\xv
\bcd
\rv_{*}
+
|\xv|
\,
r_{*}
\right)
}
\,
,
	\label{dFC_vf}
\end{eqnarray}
which, after using the relationship
\begin{eqnarray}
{\mathrm d}
\bm{B}_{\Cc}
=
-
{\bm \nabla}_{\xv}
\left(
{\mathrm d}
F_{\Cc}
\right)
\end{eqnarray}
and some vector algebra, yields Equation (\ref{dBC}).
As stated in Equation (\ref{bcC}), the obtained
$
{\mathrm d}
\bm{B}_{\Cc}
$
compensates for the photospheric radial field of the corresponding current element of the path $\Cc$.
Namely for this purpose, we used above, first, the vector $\Rv_{*}$ as a carrier for the singularity and, second, the additional multiplier ${\mathcal R}_{*}$ when deriving Equation (\ref{dFC_co}).
Both of these steps are needed to have Equation (\ref{bcC}) exactly fulfilled.

To understand the physical meaning of Equation (\ref{dFC_vf}), let us transform it to a source-centric form, similar to the one that Equation (\ref{dFCs_vfsc}) provides for the path $\Cs$.
However, the above consideration implies that it is impossible to use the path $\Cc$ as a location for the sources of the compensating potential field.
Instead, it suggests that this role belongs to another path, which is denoted by $\Css$ and obtained from $\Cc$ as a result of the inversion mapping whose point-wise definition is provided by Equation (\ref{Rvs}).
Indeed, noticing that
$
\xv
=
\rvs
+
\Rvs
$,
$
{\mathrm d}
\rv
=
-
{\mathrm d}
\Rv
$,
and
\begin{eqnarray}
{\mathrm d}
\Rvs
=
\frac
{
{\mathrm d}
\Rv
}
{
{\mathcal R}
^{2}
}
-
2
\left(
\Rv
\bcd
{\mathrm d}
\Rv
\right)
\frac
{
\Rv
}
{
{\mathcal R}
^{4}
}
\,
,
	\label{dRvs}
\end{eqnarray}
we derive from Equation (\ref{dFC_vf})
the desired source-centric expression.
\begin{eqnarray}
\left[
\frac
{\mu I}
{4\pi}
\right]
\qquad
{\mathrm d}
F_{\Cc}
=
\frac
{
{\mathcal R}_{*}
^{-1}
\:
{\mathrm d}
\Rvs
\bmt
\Rvs
\bcd
\rvs
}
{
r_{*}^2
\,
\left(
r_{*}
+
\Rvs
\bcd
\hat{\rv}_{*}
+
\left|
\rvs
+
\Rvs
\right| 
\right)
}
\,
.
	\label{dFC_vfsc}
\end{eqnarray}

The derived expression is similar to Equation (\ref{dFCs_vfsc}), except for the following two differences: first, it refers to the sources associated with the path $\Css$ rather than $\Cs$, and second, it has an additional coefficient ${\mathcal R}_{*}^{-1} \equiv {\mathcal R}$, which we call, henceforth, the modulation factor.
Except for this coefficient, the entrance of $\rvs$ and $\Rvs$ into Equation (\ref{dFC_vfsc}) is the same as that of $\rv$ and $\Rv$ in Equation (\ref{dFCs_vfsc}).
Therefore, the infinitesimal triangles that abut the path $\Css$ play a similar role: they are elementary magnetic shells whose magnetic moment surface density $\mv$ is perpendicular to the triangle planes.
However, its modulus, $|\mv|={\mathcal R}_{*}^{-1}$, is constant only along the vectors $\Rvs$, but generally changes between them.
Thus, the total magnetic potential
$
F_{\Cc}
$
is generated by the magnetic shell whose ruled surface $\Sss$ is swept out by the vector $\Rvs$ as its head moves along the path $\Css$ from the foot point $\Rv_{*1}=\Rv_{1}$ to $\Rv_{*2}=\Rv_{2}$ (see Figure \ref{f:ScSs}(a)).
 
The indicated variability of $\mv$ makes it possible to locally enhance the edge line current flowing at a given element
$
{\mathrm d}
\Rvs
$
of the path $\Css$ to a level sufficient to compensate for the boundary radial component of the coronal \BSL{} field
$
{\mathrm d}
\Bv_{I}
$.
For constant $\mv$, which implies a
constant edge current, this compensation would only be partial, as we demonstrated previously in Section \ref{ss1:mmBSl} by deriving Equation (\ref{RdBIKt}).

Note also that the variation of the edge line current along the path $\Css$ does not contradict the conservation law of the electric charge.
This variation is sustained by surface currents, which circulate in the magnetic shell $\Sss$ by flowing in or out of its edge, the path $\Css$.
In this way, the surface currents refill or drain the edge line current depending on whether the modulation factor, ${\mathcal R}_{*}^{-1}$, increases or decreases, respectively, along $\Css$ (see Figure \ref{f:ScSs}(b)).

\subsection{Entire Path $\Cc \cup \Cs$
	\label{A3}}

Let us consider how our interpretation of compensating magnetic fields in terms of magnetic shells can help simplify the \MBSL{}  defined by Equations (\ref{mmBSL}), (\ref{Thev}), (\ref{dBC}), and (\ref{dBI}).
Panels (b) and (c) in Figure \ref{f:ScSs} suggest that such a simplification is possible, because the line currents at the straight edges of $\Ss$ and $\Sss$ flow in opposite directions and reduce or even cancel each other.
For the latter to be valid, the currents must be equal in strength, which is exactly what occurs in our case.
As the modulation factor equals ${\mathcal R}_{*}^{-1}=1$ at the straight edges of $\Sss$, the local elementary magnetic shells, that is, the infinitesimal triangles adjacent to these edges, have the same magnetization for both the $\Sss$ and the $\Ss$ shells, and so the strengths of their edge line currents are the same.
This implies that a part of the first term and the entire second term in Equation (\ref{mmBSL}) cancel each other.

To perform this simplification, notice first, once again, that the field defined by \MBSL{} is independent of the shape of the path $\Cs$.
Therefore, we are free to choose this path to be identical to the curve $\Css$ and thereby to have the first shell defined on the same ruled surface $\Sss$.
Due to the linearity of the problem, we can now merge these magnetic shells into one by assigning $\mvh + \mv$ to its surface density field of the magnetic moment.
This essentially means that the resulting elementary magnetic field produced by each of the new infinitesimal magnetized triangles simply equals
$
\left(
1
-
{\mathcal R}^{-1}
\right)
{\mathrm d}
\Bv
_{\Cs}
$,
where ${\mathcal R}$ refers to points on the path $\Cs \equiv \Css$.
The modulation factor here is an algebraic sum of the previous two factors, 1 and $-{\mathcal R}^{-1}$, corresponding to the merging shells $\Ss$ and $\Sss$, respectively.
The second factor is negative, because the direction of $\mv$ is opposite to $\mvh$.
Summarizing this consideration, we finally arrive at the reduced form of \MBSL{} given by Equation (\ref{mmBSL1r}).

\section{On the Relationship between Two Methods of Decompositions}
\label{s:relation}

Our decomposition and the Gaussian decomposition of the photospheric magnetic field differ only in the way the poloidal field is treated.
Therefore, to understand the relationship between them, it is sufficient to consider only this field and its sources represented by the corresponding toroidal currents.
For simplicity, let us consider this question for a 2.5D configuration of a sheared magnetic arcade in plane geometry with Cartesian coordinates $(x,y,z)$ and the photospheric boundary at $z=0$.

\begin{figure*}[ht!]
\centering
\resizebox{0.99\textwidth}{!}{
\includegraphics{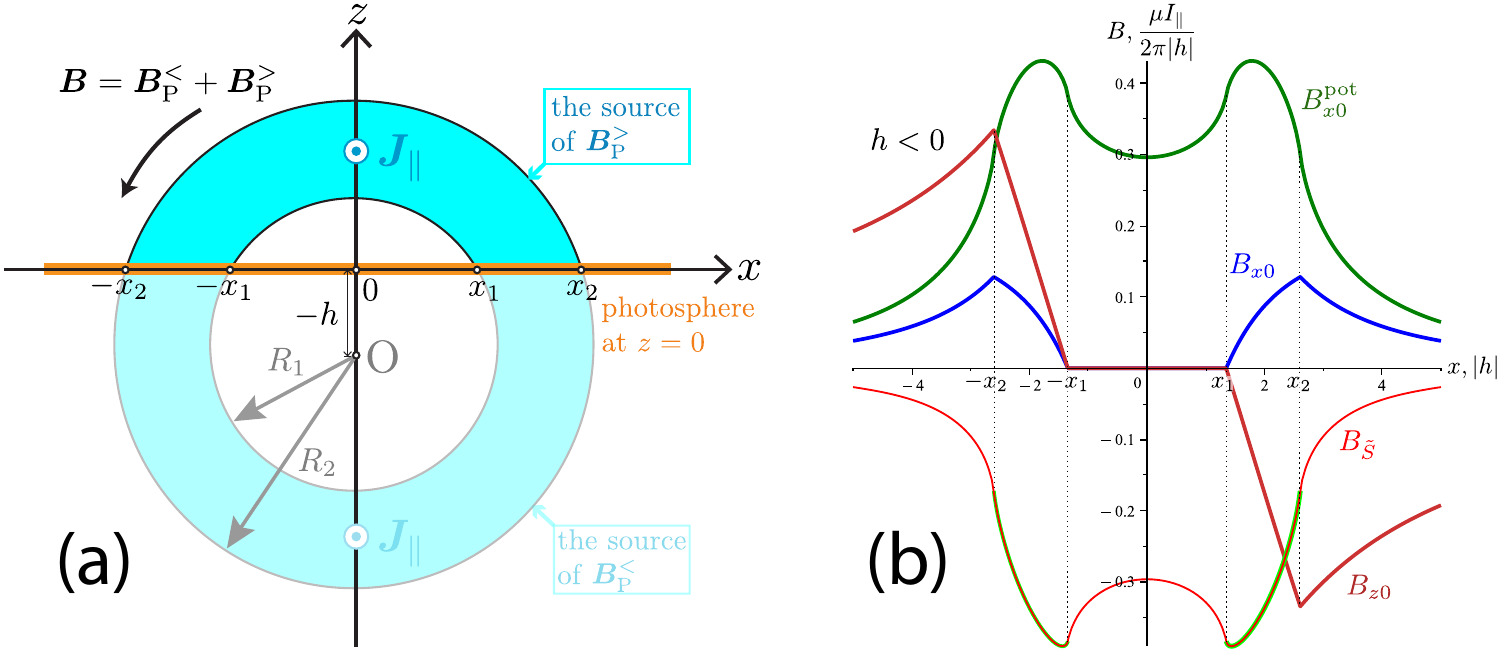}}
\caption{
A simple example of a 2.5D MFR configuration: (a) The total magnetic field is generated by an out-of-plane current with a uniform distribution of the current density $J_{\|}$ in the annulus with inner and outer radii $R_1$ and $R_2$, respectively, and the center point $\mathrm{O}$ at $(x,z)=(0,h),\; h<0$; the photospheric boundary is at $z=0$, so that the currents that flow at $z\ge0$ and $z\le0$ generate the poloidal fields $\BPgt$ and $\BPlt$, respectively.
(b) The corresponding distributions of different field components at $z=0$. 
\label{SMA_decomp_1}
}
\end{figure*}

We model this configuration using a straight hollow-core MFR with a circular cross section, where the toroidal (out-of-plane) current $I_\|$ is uniformly distributed over an annulus with inner and outer radii $R_1$ and $R_2$, respectively.
Figure \ref{SMA_decomp_1}(a) presents a perpendicular cross section of this MFR in the plane $y=0$ for the case where the MFR axis is located below the boundary at $z=h<0$ such that $|h| < R_1$.
The corresponding uniform current density is given by
%
\setcounter{equation}{0}
%
\begin{eqnarray}
J_{\|}
=
\frac
{
I_{\|}
}
{
\pi
\left(
R
^{2}
_{2}
-
R
^{2}
_{1}
\right)
}
\,
.
	\label{Jparc}
\end{eqnarray}
For such a simple axially symmetric MFR, it is not difficult to find the exact analytical expressions for the field components at the boundary, which can be written as follows:
\begin{eqnarray}
B_{x}
\est
{
z=0
}
&\equiv&
B_{x0}
(x)
=
\frac
{
\mu
I_{\|}
h
}
{
4
\pi
\left(
x^{2}
+
h^{2}
\right)
}
\left(
\frac
{
\left|
x^2
-
x_2^2
\right|
-
\left|
x^2
-
x_1^2
\right|
}
{
x_2^2
-
x_1^2
}
-
1
\right)
\,
,
	\label{Bx0u}
\\
B_{z}
\est
{
z=0
}
&\equiv&
B_{z0}
(x)
=
\frac
{x}
{h}
\,
B_{x0}
(x)
\,
,
	\label{Bz0u}
\end{eqnarray}
where $x_1=\sqrt{R_1^2-h^2}$ and $x_2=\sqrt{R_2^2-h^2}$.

Using $B_{z0}(x)$, as defined by Equations (\ref{Bx0u}) and (\ref{Bz0u}), we can derive an exact expression for the corresponding potential magnetic field in the upper half-plane, $z>0$.
This field can be represented in a compact form by a holomorphic function $\mathcal{B}^{\mathrm{pot}}(\zeta)$ of the complex variable
$
\zeta
=
x
+
i
z
$,
which itself is expressed in terms of the Schwartz integral (multiplied by $i$) as follows \citep[see, e.g.,][]{Lavrentiev1972}:
\begin{eqnarray}
\mathcal{B}
^{\mathrm{pot}}
(\zeta)
\equiv
B
^{\mathrm{pot}}
_x
-
i
B
^{\mathrm{pot}}
_z
=
\frac
{1}
{\pi}
\int
_{-\infty}
^{+\infty}
\frac
{
B_{z0}
(\xi)
}
{
\zeta
-
\xi
}
\,
\mathrm{d}
\xi
\,
.
	\label{Bx-iBz}
\end{eqnarray}
For our decomposition of the boundary field, we need only the real part of $\mathcal{B}
^{\mathrm{pot}}
(x+i\, 0)$, which we can calculate from Equation (\ref{Bx-iBz}) to obtain
\begin{eqnarray}
B
^{\mathrm{pot}}
_{x}
\est
{z=+0}
&\equiv&
B
^{\mathrm{pot}}
_{x0}
(x)
=
\frac
{1}
{\pi}
\Re
\left(
\lim
_{z=+0}
\int
_{-\infty}
^{+\infty}
\frac
{
B_{z0}
(\xi)
}
{
x
+
i
z
-
\xi
}
\,
\mathrm{d}
\xi
\right)
\nonumber
\\
&=&
\frac
{
\mu
I_{\|}
}
{
\pi^2
\left(
x_2
+
x_1
\right)
}
\left\{
1
+
\frac
{h}
{
x_2
-
x_1
}
\left[
\frac
{
x_{2}^{2}
+
h^{2}
}
{
x^2
+
h^2
}
\arctan
\left(
\frac
{h}
{x_2}
\right)
-
\frac
{
x_{1}^{2}
+
h^{2}
}
{
x^2
+
h^2
}
\arctan
\left(
\frac
{h}
{x_1}
\right)
\right.
\right.
\nonumber
\\
&&
\qquad
\qquad
\qquad
+
\left.
\left.
\frac
{x}
{2h}
\left(
\frac
{
x^2
-
x^2_{1}
}
{
x^2
+
h^2
}
\ln
\left|
\frac
{
x
+
x_1
}
{
x
-
x_{1}
}
\right|
-
\frac
{
x^2
-x^2_{2}
}
{
x^2
+
h^2
}
\ln
\left|
\frac
{
x
+
x_2
}
{
x
-
x_2
}
\right|
\right)
\right]
\right\}
\,
.
	\label{Bpot0}
\end{eqnarray}
Our poloidal magnetic field at the boundary, $B_{\St}$, is then simply determined as
\begin{eqnarray}
B_{\St}
=
B_{x0}
-
B
^{\mathrm{pot}}
_{x0}\,
,
	\label{BSt}
\end{eqnarray}
the explicit form of which can be obtained by substituting Equations (\ref{Bx0u}) and (\ref{Bpot0}) in this expression.
Panel (b) in Figure \ref{SMA_decomp_1}
presents graphs of $B_{z0}$, $B_{x0}$, $B^{\mathrm{pot}}_{x0}$, and $B_\St$ as functions of $x/|h|$ for particular values of $R_1/|h|$ and $R_2/|h|$ that correspond to the configuration depicted in panel (a).

\begin{figure*}[ht!]
\centering
\resizebox{0.99\textwidth}{!}{
\includegraphics{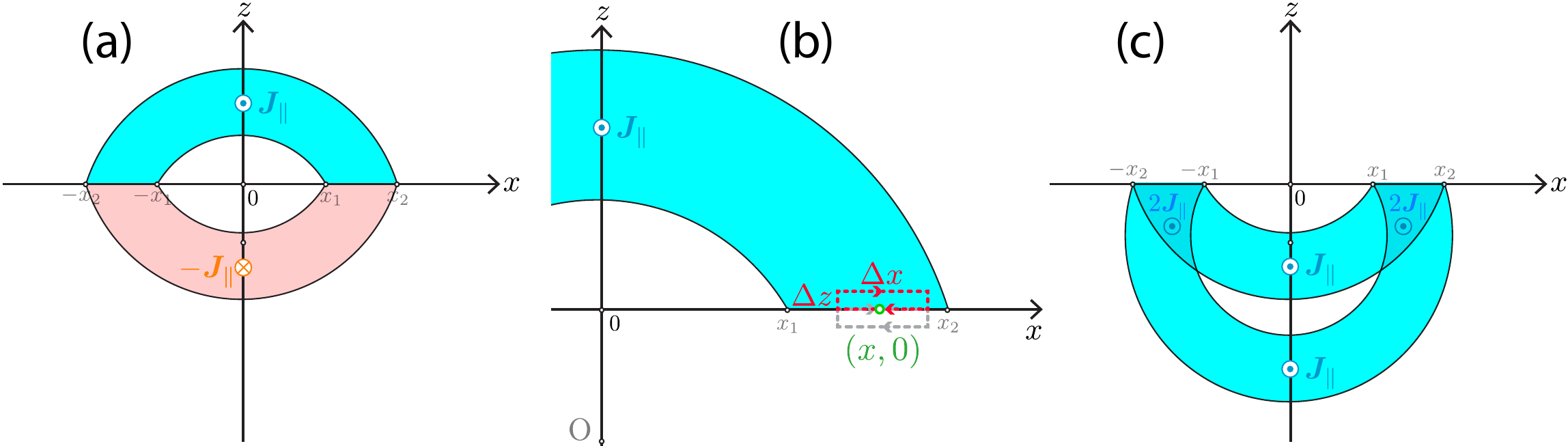}}
\caption{
(a) The coronal (cyan) and mirror (pink) currents that produce the magnetic field $\Btld$ such that $\tilde{B}_z(x,0)=0$ and $\tilde{B}_x(x,0)=B_\St(x)$.
(b) The Stokesian loops (red and gray dashed rectangles) contracting to a given point $(x,0)$ at the boundary. 
These are used to derive Equation (\ref{BSt2BPxgt}).
(c) The current sources of the potential magnetic field that is determined from $B_{z0}(x)$ by Equations (\ref{Bx0u})\textendash(\ref{Bx-iBz}).
\label{SMA_decomp_2}
}
\end{figure*}

In the configuration under study, the compensating magnetic field is produced by mirror currents, so that the total current-carrying field $\Btld$ with $\ez \bcd \Btld(x,0)=0$ is generated by a system of currents whose distribution is antisymmetric about $z=0$, as shown in Figure \ref{SMA_decomp_2}(a).
Due to this symmetry and the potentiality of the poloidal field $\BPgt$ outside the annulus, the corresponding magnetic field at the boundary in this region is given by\footnote{We thankfully acknowledge Mark Linton for deriving and providing us with Equations (\ref{BSt2BPxgt}) and (\ref{Bx0pot}) under the implicit assumption that  
$
\ex
\bcd
\BPlt
(x,+0)
\equiv
\ex
\bcd
\BPgt
(x,-0)
$.
}
\begin{eqnarray}
B_\St
=
2
\,
\ex
\bcd
\BPgt
\est
{z=+0}
=
2
\,
\ex
\bcd
\BPgt
\est
{z=0}
=
2
\,
\ex
\bcd
\BPgt
\est
{z=-0}
\,
.
	\label{BSt2BPxgt}
\end{eqnarray}
It is less obvious that this relationship also holds in the annulus region itself, where the current density $J_\|\ne0$.
Consider a rectangular loop
 of width $\Delta x = k \epsilon$ and height $\Delta z = \epsilon$ (red dashed rectangle with an aspect ratio $k>1$ in Figure \ref{SMA_decomp_2}(b)), one side of which is centered at a given point $(x,0)$.
Applying the Stokes theorem \citep[see, e.g.,][]{Jackson1962} to this loop, we obtain
$
\ex
\bcd
\BPgt
(x,+0)
-
\ex
\bcd
\BPgt
(x,0)
=
0
$
in the limit $\epsilon \rightarrow 0$.
This is because the circulation of
$
\BPgt
$
and the total current that flows through the loop vanish as $\epsilon$ and $\epsilon^2$, respectively.
Using the same approach for a similar loop that is mirrored across the boundary surface (gray dashed rectangle in Figure \ref{SMA_decomp_2}(b)), we obtain 
$
\ex
\bcd
\BPgt
(x,0)
-
\ex
\bcd
\BPgt
(x,-0)
=
0
$.

These equalities confirm that Equation (\ref{BSt2BPxgt}) indeed remains valid in the current-carrying region.
It is also clear that the same conclusion is valid for an arbitrary continuous distribution of the current density at $z\ge0$ and even for distributions that have an integrable singularity $\sim z^{-\nu}$, $0<\nu<1$, at the boundary.
However, if the singularity behaves as the Dirac delta function, $\delta(z\!-\!0)$, (or in other words, there is a current sheet at $z=+0$) both the circulation of
$
\BPgt
$
and the total current that flows through the red loop vanish at the same rate $\sim \epsilon$.
In this case, we will obtain a finite jump of 
$
\ex
\bcd
\BPgt
$
that is proportional to a local value of the surface current density  \citep[see, e.g., Equation (I.20) in][]{Jackson1962},
which implies that Equation (\ref{BSt2BPxgt}) is no longer valid.
Further analysis of this case is then required, which is, however, beyond the scope of the present article.

Equation (\ref{BSt2BPxgt}) has been first verified numerically by using a Cartesian version of CICCI on the magnetic field in the $z=0$ plane (Mark Linton; personal communication). 
Later, we also checked this relationship by integrating (both numerically and analytically) the contributions of elementary currents $J_{\|}\, \dx\, \dz$ over the upper part of the annulus at $z>0$.
The red curve representing $B_{\St}$ in Figure \ref{SMA_decomp_1}(b) is perfectly aligned with the thick green curve that was obtained by direct integration, so the former is not visible. 
This shows that a jump in the toroidal current density across the boundary causes {\it only a jump in the normal derivatives of} $\BPgt$, {\it but not in their values}.
This important conclusion is based on the local properties of $\BPgt$ and is consistent with the above consideration of Stokesian loops, so it must be valid even for general 3D configurations with planar or spherical boundaries.

The relationship between the tangential components of our potential field and the Gaussian poloidal fields can be found as well.
To this end, note that the residual of the current configurations, which are shown in  Figures \ref{SMA_decomp_1}(a) and \ref{SMA_decomp_2}(a), provides the current configuration presented in Figure \ref{SMA_decomp_2}(c), which, by construction, produces
\begin{eqnarray}
B
^{\mathrm{pot}}
_{x0}
=
\left(
\ex
\bcd
\BPlt
-
\ex
\bcd
\BPgt
\right)
\est{z=+0}
\,
.
	\label{Bx0pot}
\end{eqnarray}

The obtained Equations (\ref{BSt2BPxgt}) and (\ref{Bx0pot}) show how our decomposition and the Gaussian decomposition of the tangential components of the photospheric poloidal field are related to each other in the 2.5D configuration considered.
In fact, we have recently derived similar relationships between the two decompositions in arbitrary 3D configurations by representing these components in terms of Green functions, which will be described in a forthcoming publication.

The considered example additionally shows that for magnetic configurations with $\Jv$ distributions {\it without} current sheets at the solar surface, only the Gaussian decomposition provides the authentic separation of the magnetic fields generated by coronal and subsurface currents, respectively.
This is because our decomposition method generally implies the presence of fictitious subsurface mirror currents that compensate for the normal component of the surface field produced by the coronal currents.

On the other hand, with respect to coronal configurations, our mirror currents can be interpreted as a replacement for surface current sheets.
The latter can be formed at the surface ($z=0$ in Cartesian geometry or $r/R_\sun=1$ in spherical geometry) because of the line-tying effect, which shields the conductive and dense solar interior from the magnetic flux of evolving coronal currents (see Section \ref{ss:preface}).
Then, we can interpret $B_\St$ given by Equation (\ref{BSt2BPxgt}) as the sum of two identical poloidal fields, which are produced by all coronal currents and the corresponding shielding surface current, respectively.
Similarly, Equation (\ref{Bx0pot}) provides the difference between two tangential poloidal fields generated by all subsurface currents and the shielding surface current, respectively.
In this interpretation, the negative term in Equation (\ref{Bx0pot}) corresponds exactly to the field component of the surface current sheet on its lower side $z=-0$.

Hence, compared to the Gaussian method, which implies a free penetration of the coronal flux into the solar interior, our method relies on the opposite idealization where the interior remains perfectly shielded from this flux.
In reality, of course, neither of these two extremes is realized on the Sun.
As discussed in Section \ref{ss:preface}, significant shielding can be expected, for example, during the dynamic phase of solar eruptions, but its degree likely depends on how rapidly the current evolves in the corona.
It is also clear that the shielding current is concentrated in thin layers rather than in idealized current sheets of zero width, since the real corona and photosphere are imperfect conductors.
In this respect, it would be worthwhile to develop an approach that combines the two methods, in order to determine the correct decomposition of the magnetic field on the upper side of such current layers for the more realistic condition of partial shielding of the solar interior.

\bibliography{MBSL}{}
\bibliographystyle{aasjournalv7}

%

\end{document}